\documentclass[referee]{raa}           
\usepackage{graphicx,times}
\usepackage{url}
\usepackage{natbib}
\usepackage{lscape}
\usepackage{threeparttable}

\begin{document}
\title{Candidate members of the Pal 5, GD-1, Cetus Polar, and Orphan tidal stellar halo streams from SDSS DR9, LAMOST DR3 and APOGEE catalogues}

 \volnopage{ {\bf 201x} Vol.\ {\bf X} No. {\bf XX}, 000--000}
   \setcounter{page}{1}

   \author{Guang-Wei Li\inst{1} \and Brian Yanny\inst{2} \and Hao-Tong Zhang\inst{1}  
       \and Zong-rui Bai\inst{1} \and Yue Wu\inst{1} \and Yi-qiao Dong\inst{1} \and Ya-juan Lei\inst{1} 
       \and Hai-long Yuan\inst{1}   
   \and Yong-Hui Hou\inst{3} \and Yue-Fei Wang\inst{3} \and Yong Zhang\inst{3}}

  \institute{Key Lab for Optical Astronomy, National Astronomical Observatories, Chinese Academy of
 Sciences, Beijing 100012, China; {\it lgw@bao.ac.cn}\\
         \and
                Fermi National Accelerator Laboratory, Batavia, IL 60510, USA\\
                \and  
                    Nanjing Institute of Astronomical Optics \& Technology, National Astronomical Observatories, Chinese Academy of Sciences, Nanjing 210042, China \\
}	

 \date{Received~~201x month day; accepted~~201x~~month day}

\abstract{
We present candidate members of the Pal 5, GD-1, Cetus Polar, and Orphan tidal stellar streams
found in LAMOST DR3, SDSS DR9 and APOGEE catalogs. In LAMOST DR3, we find 20, 4, 24 high confidence 
candidates of tidal streams GD-1,  Cetus Polar and Orphan respectively. We also list from the
SDSS DR9 spectroscopic catalog 59, 118, 10 high confidence candidates of tidal streams 
Cetus Polar, Orphan and Pal 5, respectively.  Furthermore, we find 7 high confidence candidates of the 
Pal 5 tidal stream in APOGEE data. Compared with SDSS, the new candidates from LAMOST DR3 are brighter, so
that together, more of the color-magnitude diagram, including the giant branch can be explored.
Analysis of SDSS data shows that there are 3 metallicity peaks of the Orphan stream and 
also shows some spatial separation. LAMOST data confirms multiple metallicities in this stream.
The metallicity, given by the higher resolution APOGEE instrument, of the Pal 5 tidal stream is
 $\rm [Fe/H] \sim -1.2$, higher than that given earlier by SDSS spectra.  
Many previously unidentified stream members are tabulated here for the first time, 
along with existing members, allowing future researchers to further constrain the orbits of these 
objects as they move within the Galaxy's dark matter potential.
}

\authorrunning{G.-W. Li et al. }            
   \titlerunning{Candidates of streams}  
   \maketitle

%

\section{Introduction}
There are two popular models for the how our Galaxy formed. The first, given by \cite{egg62}, suggested
that the Galaxy was born from a single rapid collapse of a massive cloud of gas. 
Later, \cite{sea78} suggested that an inner halo may have come from a large early collapse, 
but the outer halo independently evolved over a much longer period of time, 
and during this time, many small stellar systems merged into the halo, and were tidally disrupted by the 
Galaxy's potential. The standard $\Lambda$CDM cosmological model also favors big galaxies
growing from the merger of smaller units.
\par

The largest and most famous Milky Way halo stellar stream is that associated with the Sagittarius dwarf 
galaxy \citep{iba94}, mapped out in 2MASS giants by \cite{maj03}. Much progress in stream detection 
occurred with the release of the SDSS dataset \citep{yor00}.  Four streams: GD-1, Orphan, 
Cetus Polar Stream (CPS), and the Pal 5 tidal stream were discovered using SDSS data. GD-1 is a $63^{\circ}$ 
narrow stream, found by \cite{gri06}. Later, \cite{wil09} fit an orbit.  \cite{yan09a}
noticed that there is a tidal stream near the Sgr trailing tidal tail which was named the CPS by \cite{new09}. 
Its parameters were given by \cite{yam13} using SDSS DR8.  The Orphan stream was found by \cite{gri06} 
and \cite{bel06}, but its orbital parameters were not clear until \cite{new10} gave them using SDSS DR 7. 
Pal 5 is a globular cluster which is  being disrupted. Its long tail was firstly discovered by  \cite{ode01},  
which spans more than $23^{\circ}$ \citep{car12}, and is the most obvious stream associated with a 
Galactic globular cluster. 
\par

Having an accurate census of stream members, as well as their spectroscopic properties (including radial 
velocity, and parameters which may be used to estimate absolute magnitude -- and thus distance) is
crucial to determining accurate orbits of the streams.  In turn, having an accurate orbit for a set of streams 
allows us to probe the (dark matter dominated) gravitational potential at a variety of distances and 
directions throughout the Milky Way's halo \citep{new10}.
\par

A major unresolved question in Galactic dynamics is understanding in detail the shape (i.e. oblate, 
prolate, spherical, lumpy, changing-with-radius?) and extent (total mass and drop-off with radius) 
of our Galaxy's dark matter potential and the dark halo's shape and size.  For instance,
is the halo triaxial in nature as suggested by \cite{lmj09}?  Having extensive, accurate stellar stream 
membership information, along with radial velocity and photometric parallax information for member 
stars can help resolve this important question.  This work adds to our list of known stream members, 
with spectroscopic velocity and other stellar parameters for four halo streams.

\par
The Large Sky Area Multi-Object Fiber Spectroscopic Telescope 
(LAMOST, also called the Guo Shou Jing Telescope) \citep{cui12, wang96, su04} is a special reflecting Schmidt telescope
with field of view (FOV) $5^\circ$ and effective aperture 3.6m - 4.9 m. There are 4,000 fibers on its focal 
plane and they can record 4,000  spectra at once.  Its wavelength coverage is 365 nm - 900 nm
with R $\sim$ 1,800. Each of 16 spectrographs records images of 250 fibers on a  4136 pixel $\times$ 4160 pixel CCD. As of May 30, 2015, more than 3 million A, F, G and K stellar spectra with parameters have been released 
\citep{luo12, zhao12}(see \url{http://dr3.lamost.org/}). 
\par

In this paper, we search for and describe parameters for high confidence stellar candidates 
of the GD-1, Orphan, CPS and Pal 5 tidal stream in the spectral data of LAMOST DR3, SDSS DR9 \citep{yan09b}
and APOGEE \citep{maj15}. 
\par

\section{Candidates of Streams}  \label{sect2}
We search for stream members, primarily giants ($0< \rm log\> g < 3.5$), in the LAMOST DR3, SDSS DR9 
APOGEE spectral databases.  As is common in the literature, magnitudes with subscript 0 indicate they have 
been corrected by the extinction given by \cite{sch98}, (not the more recent \cite{sch11} -- the differences 
are tiny at these higher Galactic $|b|$).  All $g$ and $r$ band magnitudes in this paper are from SDSS DR9. 
Because the [Fe/H] values estimated by the standard LAMOST processing pipeline \citep{wu14} have a lower limit of $-2.4$, 
there is no star with quoted [Fe/H] less than $-2.4$ in LAMOST DR3.  For SDSS stars with spectra, \cite{new09} found the quoted 
FEHWBG \citep{wil99} (WBG) parameter is a better measure of metallicity than FEHADOP for blue BHBs, 
so we use WBG metallicity for stars with $(g-r)_0 < 0.2$ and FEHADOP (adopted) metallicity for stars with 
$(g-r)_0>0.2$ . For a star observed many times, we only retain the spectrum with the highest Signal-to-noise 
ratio (SNR) in the $g$ and $r$ bands for LAMOST spectra, and that with the highest SNR for SDSS and APOGEE.  
We convert radial velocities to the Galactic standard of rest velocities ($V_{gsr}$) using the formula  $V_{gsr} = RV 
+ 10.1\cos b \cos l + 224 \cos b \sin l + 6.7 \sin b$, where $RV$ is the heliocentric radial velocity in km s$^{-1}$, 
while $(l, b)$ are the Galactic coordinates of the star. 

For each of the four streams accessible with LAMOST or SDSS or APOGEE spectroscopy, and which have SDSS 
g \& r-band photometry, we search in the up to seven-dimensional space of: 1,2) (l,b) position along the orbit; 3) distance 
(determined by photometric parallax using the cataloged star's color, magnitude and spectral type (giant or 
dwarf) depending on surface gravity);  5) velocity and (lack of measureable) 6,7) proper motion to select candidates 
and give confidence estimates of candidates' membership in a given stream. We give each candidate a number
(1, 2 or 3) to describe our confidence in stream membership, with a higher number indicating lower confidence.

\subsection{Candidates of the GD-1 Stream}
\label{sec-gd1}
The study of \cite{wil09} has given all the GD-1 stream candidates present in SDSS data, so here we only search 
for GD-1 candidates in LAMOST DR3, using the same method as \cite{wil09}.  All giant candidates match the 
following criteria:
\par
1) The GD-1 positional locus is $\delta = -864.5161 + 13.22518\alpha   - 0.06325544 \alpha^2 + 0.0001009792 \alpha^3$.  
 Candidates should be within $\delta \pm 1^{\circ}$.
\par
2) There are 7 set (regions) of stars close in position and velocity listed in \cite{wil09}. Regions 1, 4, 5, 6 have high
confidence stream members. Between Region 1 and Region 4, Region 2 is more reliable than Region 3 (in terms 
of groupings of velocities and color-magnitude properties of candidate stars), so we only use data of Region 2. 
Beyond Region 6, the only data we can use is Region 7 though it is not so reliable as Region 1, 4, 5, 6. Thus, we use  
$V_{gsr}$ in Region 1, 2, 4, 5, 6, 7 to select candidates. We generate the velocity trendline by interpolation and 
extrapolation of velocities of these regions, then select candidates within 30 km s$^{-1}$ around the trendline. 
\par
3) Galactic proper motions and errors are calculated from equatorial proper motions with errors in the SDSS DR9 
(based on the UCAC4 catalog).  We select high confidence candidates by limiting their proper motion within
 2$\sigma$ of the expected proper motions given by \cite{wil09}.
\par
4) Metallicities [Fe/H] should be in the range $[-2.5,-1.5]$. 
\par

\begin{figure}[h]
  \centering
   \includegraphics[width=160mm,height=100mm]{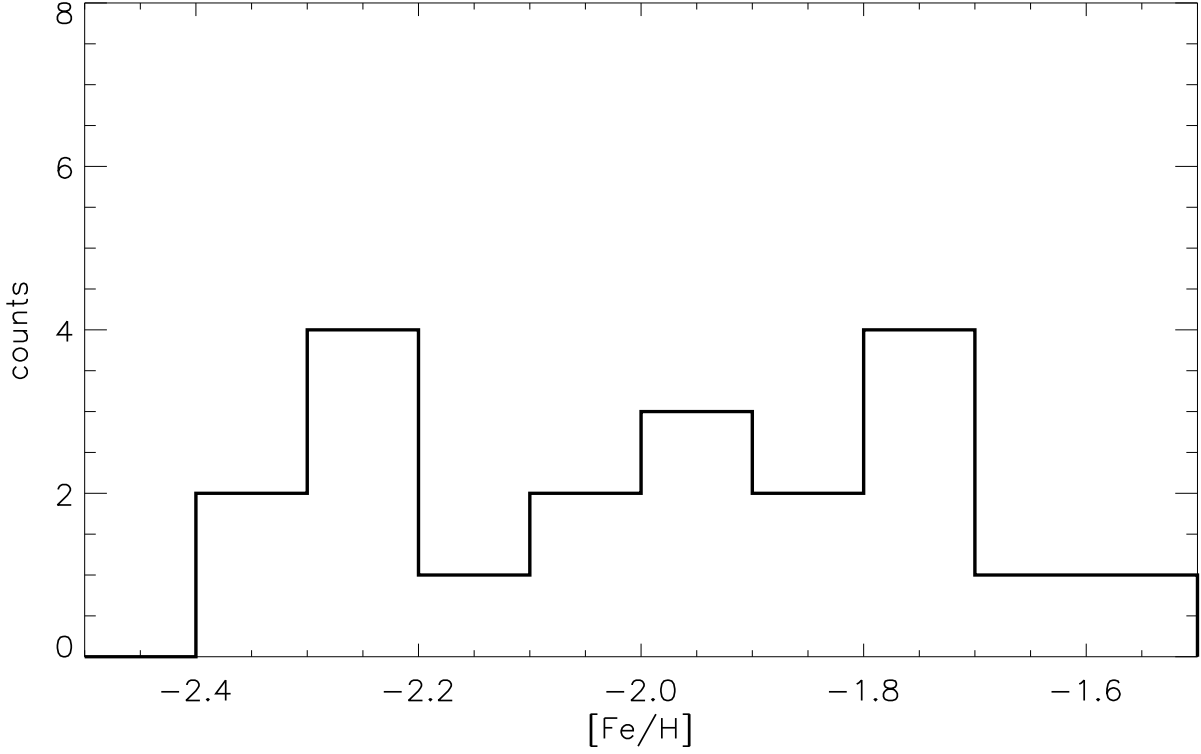}

  \caption{The metallicity distribution of GD-1 stars in LAMOST DR3 that match all criteria. }

  \label{gd1-feh}
\end{figure}

Fig. \ref{gd1-feh} shows their metallicity distribution.  As shown by Fig. 2 in \cite{gao15}, LAMOST overestimates 
metallicities for the most metal-poor stars, and their variance is larger than metal-richer stars. The metallicity 
distribution of GD-1 member candidates from SDSS DR7, shown in Fig. 5 in \cite{wil09}, is also broad. Thus, although these
stars span broad metallicity range, we still consider them candidate members of the GD-1 stream.

\begin{figure}[h]
  \centering
   \includegraphics[width=160mm,height=100mm]{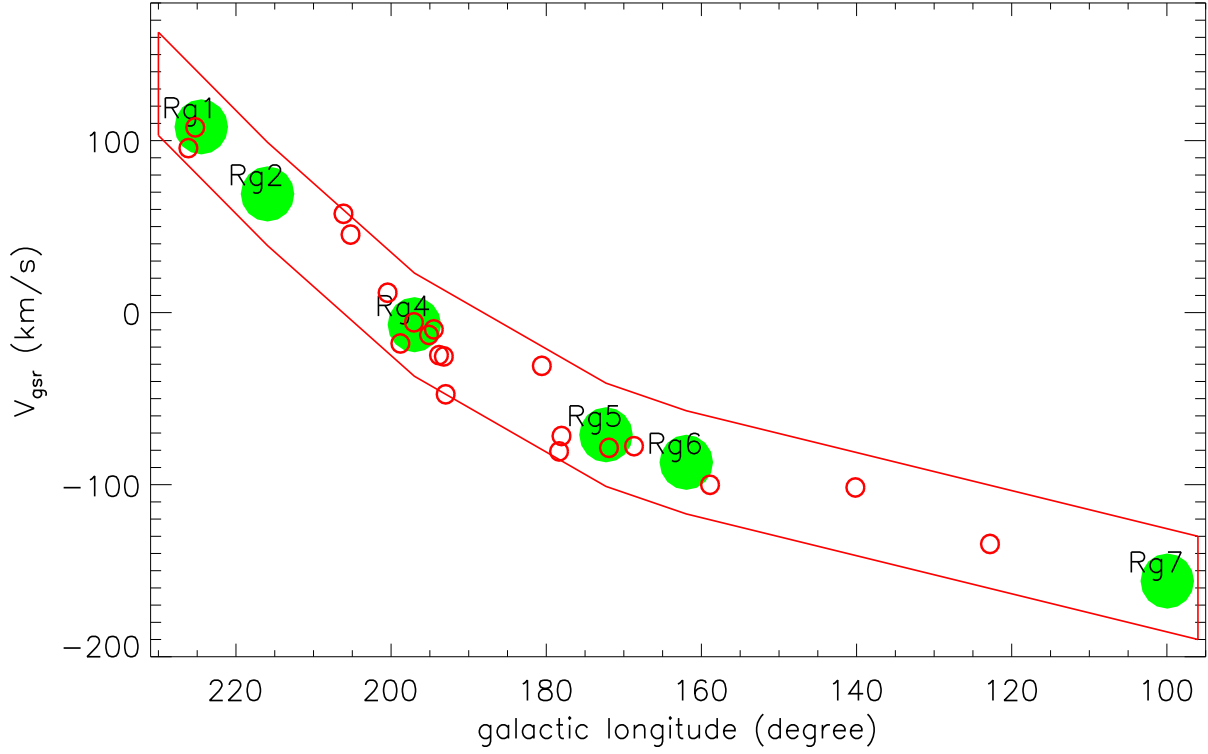}

  \caption{Big green filled circles are $V_{gsr}$ of regions given by \cite{wil09}. The red polygon is the area where 
  we select candidates of GD-1. The red circles are the high  confidence candidates in LAMOST DR3.}
  \label{gd1-vgsr}
\end{figure}

\begin{figure}[h]
  \centering
   \includegraphics[width=160mm,height=100mm]{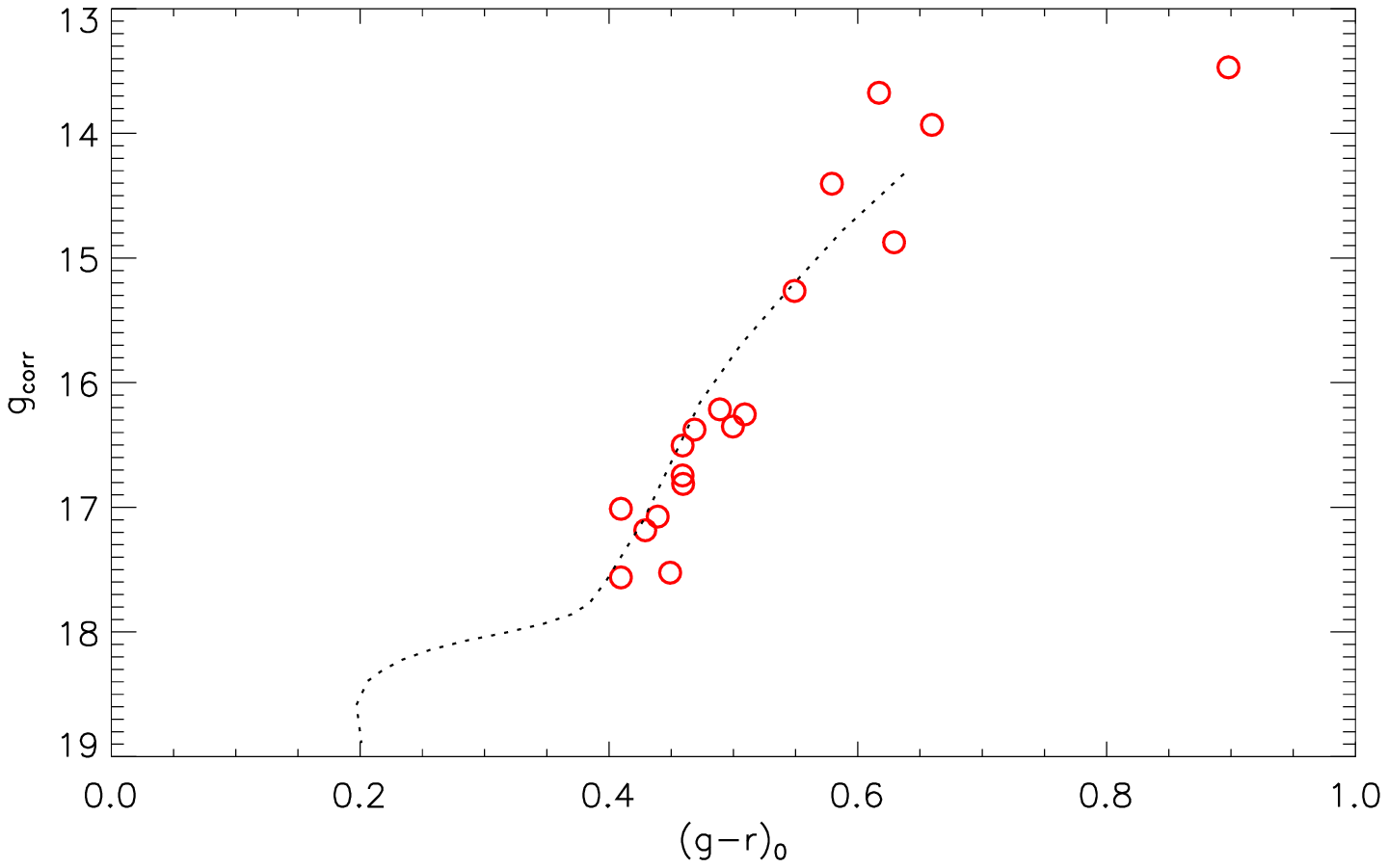}

  \caption{The red circles are the high confidence candidates of GD-1. The background line is the M92 fiducial locus. 
  Each candidate is corrected to the distance of M92, which is 8.2kpc. Magnitudes of the top 5 stars are inaccurate due to saturation, and
  one star has no SDSS photometry -- it is not shown in the figure. }

  \label{gd1-cmd}
\end{figure}

The red polygon in Fig. \ref{gd1-vgsr} shows the area where the high confidence candidates rest.  Their color-magnitude 
diagram(CMD, hereafter) is shown in Fig. \ref{gd1-cmd}.  We estimate their absolute magnitudes as they are giants similar 
to metal poor giants in the globular cluster M92, using the M92 isochrone and then remove candidate stars with implied 
distances far from the GD-1 stream distance.  This leaves us with 20 high confidence candidates. Magnitudes of the brightest 
five stars, which we believe are GD-1 members, are not reliable because SDSS CCD saturates around $r \sim 14.5$.  
For a stream, brighter members are fewer, so if these five bright stars is really belong to GD-1 stream, they will greatly help 
us to understand the stream.

One star in Region 1 has no SDSS photometry, but we still retain it, because there are only two candidates in Region 1,
if it is really a GD-1 member, it would be valuable to study this Region.

The confidence level of each candidate is given based on its position on Fig. \ref{gd1-vgsr} and  Fig. \ref{gd1-cmd}.  
Let $Dv_i$ denote the difference between $V_{gsr}$ of the $i$th candidate and the trendline at its longitude, and $\sigma^2_1
 = \sum_{i\in\phi}Dv^2_i/(n-1)$, where $\phi =\{i | |Dv_i| < 3 \sigma_1 \}$ and $n$ is the 
number of the elements in $\phi$; Let $Dc_i$ denote the difference between 
 $(g-r)_0$ of the $i$th candidate and the M92 isochrone  at its $g_{corr}$, and $\sigma^2_2
 = \sum_{i\in\psi}Dc^2_i/(m-1)$, where $\psi =\{i | |Dc_i| < 3 \sigma_2 \}$ and $m$ is the 
number of the elements in $\phi$. Then we 
calculate the $f_i$ for the $i$th candidate by the formula:
$f_i = (\frac{Dv_i}{\sigma_1})^2 + (\frac{Dc_i}{\sigma_2})^2$.
If $f_i \leq 2$, then the confidence level of the $i$th candidate is set  to be 1; if $ 2< f_i \leq 10$, then 
the confidence level is set to be 2; if $ f_i > 10$, then the confidence level is set to be 3.

\par
We searched for GD-1 candidates in APOGEE spectral data, but found none.

\subsection{Candidates of the Cetus Polar Stream}
We select CPS candidates by the method given by \cite{yam13} .  All giant candidates should match the following 
criteria:
\par
1)  Metallicities should be in  $[-2.5,  -1.5]$;
\par
2) Distances to the Galactic great circle $l = 143^\circ$ should be less than $15^\circ$;
\par
3) We find that the the stream velocity formula $V_{gsr} = -41.67-(0.84\times b)-(0.014\times b^2)$  given by 
\cite{yam13} has a typographical error,  and the correct formula is $V_{gsr} = -41.67+(0.84\times b)+(0.014\times b^2)$, and
we use this. We select candidates within $V_{sgr} \pm 20$ km s$^{-1}$.
 \par
4) We use the fiducial sequence of NGC 5466 as a reference to select confidence candidates in the CMD, 
then use equation 2 in \cite{yam13} to calculate their corrected magnitudes $g_{corr}$. All 
stars should be  within $-0.01800g_{corr}^3 + 0.98473g_{corr}^2 - 18.05165g_{corr}+111.43819 - 0.04 < (g-r)_0 < 
-0.01800g_{corr}^3 + 0.98473g_{corr}^2 - 18.05165g_{corr}+111.43819 + 0.04$ and $16 < g_{corr} < 18.5$. 
\par
5) Proper motions in R.A. and Dec. are all less than 6 mas/yr.
\par

\begin{figure}[h]
  \centering
   \includegraphics[width=160mm,height=100mm]{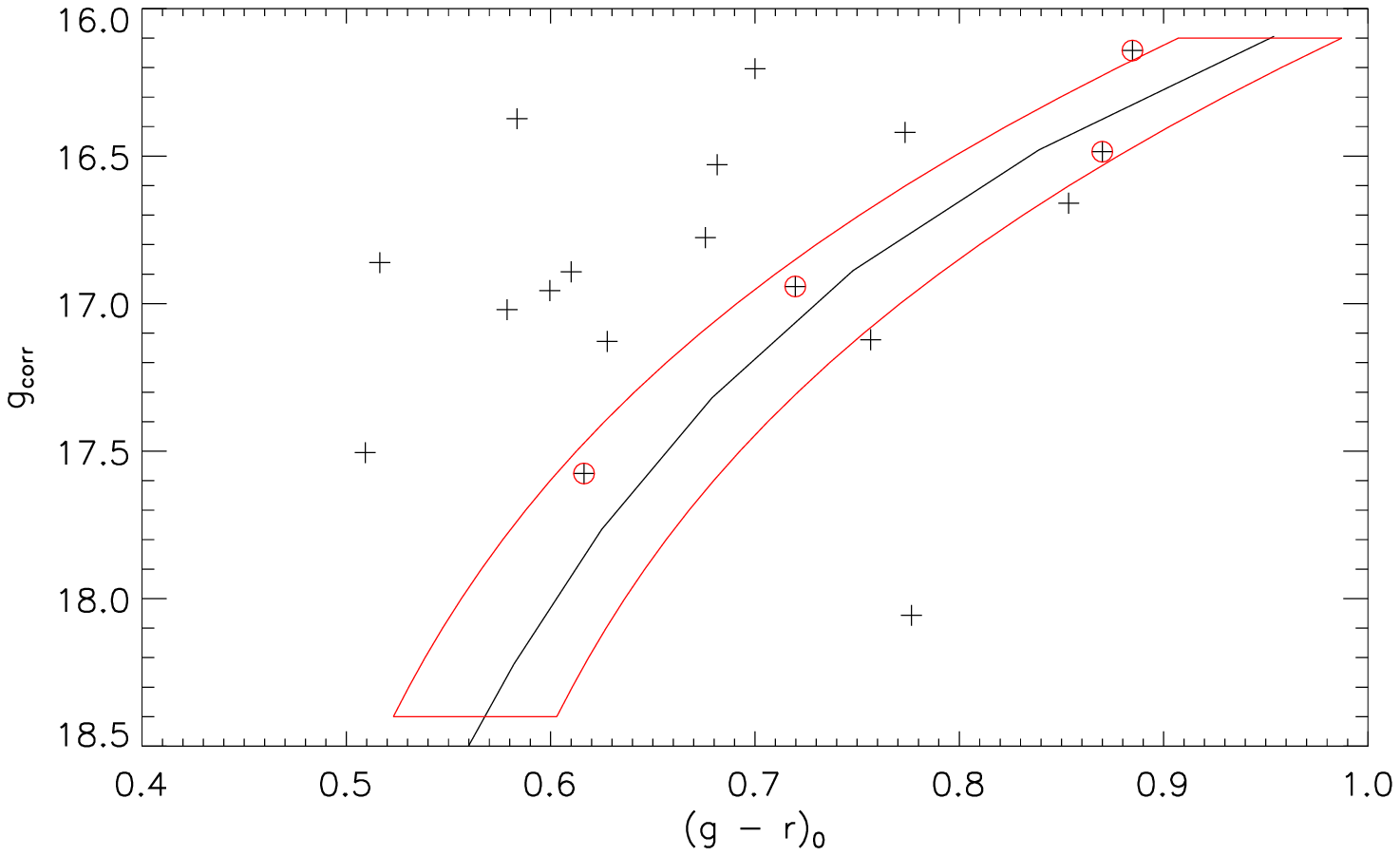}

  \caption{This figure shows the CPS candidates in LAMOST DR3. The 
  stars denoted by crosses are stars which match criteria 1, 2, 3, 5, while the high confidence candidates are denoted
  by the red circles. The black line is the  fiducial sequence of NGC 5466  shifted to the distance modulus 17.389.  
  The area given by criteria 4 is shown by red lines.
  }

  \label{cetus-cmd}
\end{figure}

\begin{figure}[h]
  \centering
   \includegraphics[width=160mm,height=100mm]{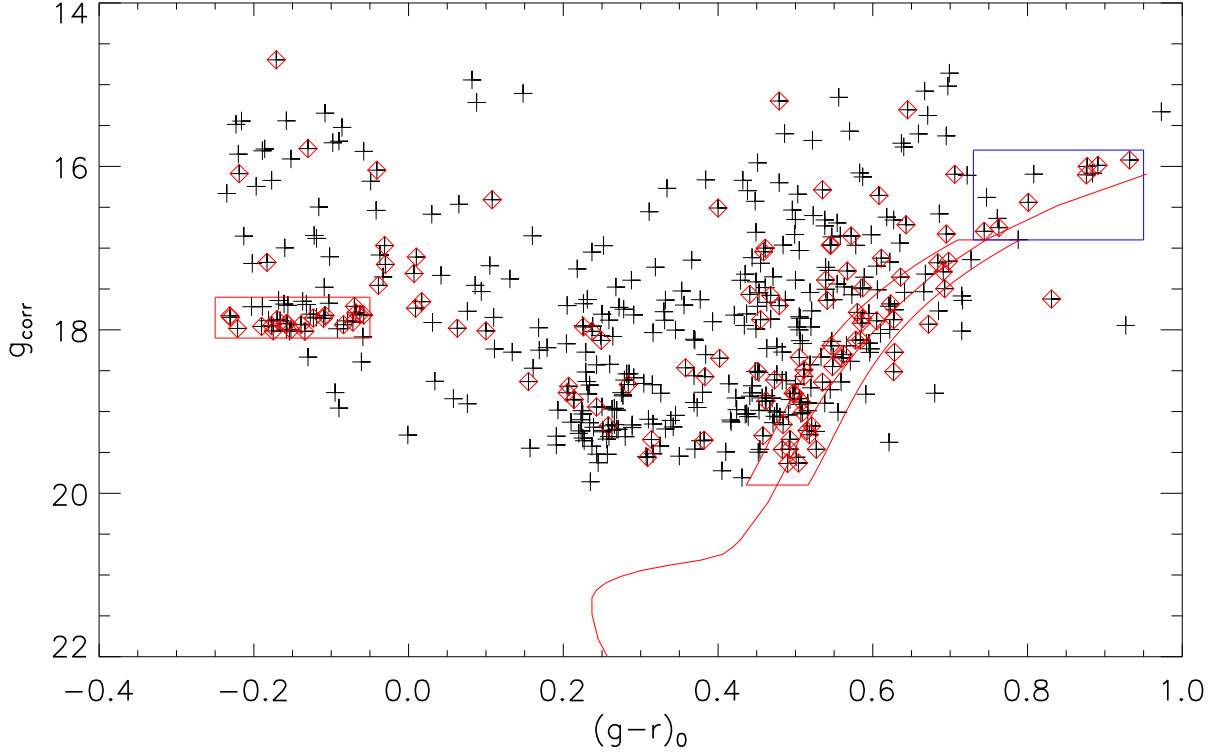}

  \caption{The crosses are the candidates of the CPS in SDSS DR9 that match the criteria 2 - 5. The red diamonds are stars 
  with metallicity in [$-2.5$, $-2.0$]. The area enclosed by red and blue lines is where we select candidates. The 
  blue rectangle is the new area not given by \cite{yam13}.}

  \label{cetus-sdss-cmd}
\end{figure}

Fig. \ref{cetus-cmd} shows the CMD of these candidates in LAMOST DR3. The red circles are candidates we select.
The stars denoted by crosses are candidates match criteria 1, 2, 3, 5. From Fig. \ref{cetus-sdss-cmd} we can see that 
the colors of brighter Orphan stars go little upward from  the fiducial sequence of NGC 5466, so we  retain the top 
right corner star. 
\par

\begin{figure}[h]
  \centering
   \includegraphics[width=160mm,height=100mm]{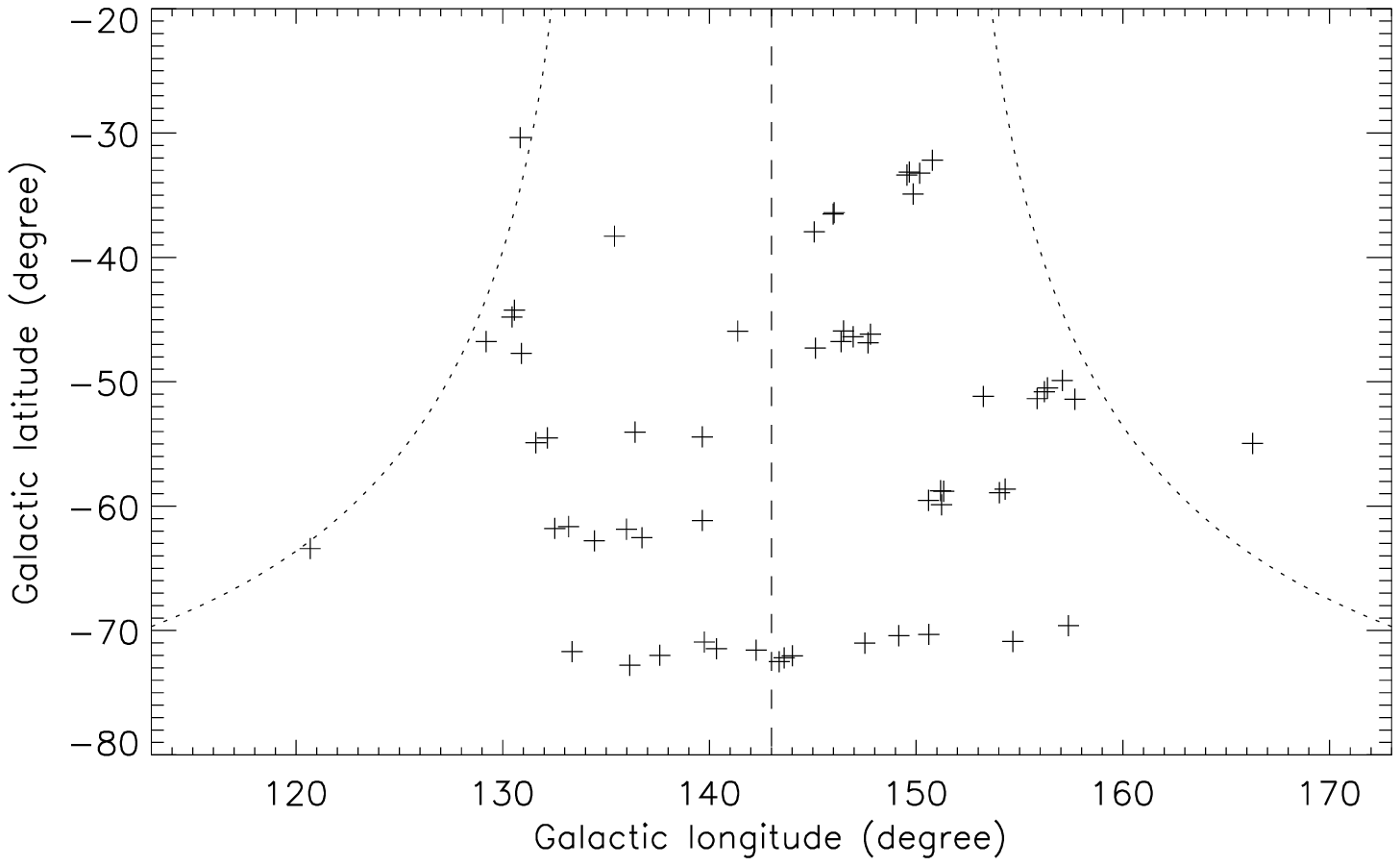}

  \caption{This figure shows the distribution range of the CPS in Galactic coordinates. The central line is $l = 143^{\circ}$, 
  the two dotted lines are the bounds $10^{\circ}$ from the central line on the celestial sphere, while the crosses 
  are the candidates in SDSS DR9. }

  \label{cetus-sdss-pos}
\end{figure}

Additionally, we search for additional CPS candidates within SDSS DR9 by the same criteria as the above except 
the metallicity is restricted to $[-2.5, -2]$. Here obtain 59 candidates, including 21 BHBs. Fig. \ref{cetus-sdss-cmd} 
shows the CMD of stars in CPS area. The red diamonds in the area enclosed by red and blue lines are the high 
confidence candidates.  While many of these objects were already shown in \cite{yam13}, there is, at red colors,
a blue rectangle containing new candidates not searched in \cite{yam13}. Fig. \ref{cetus-sdss-pos} shows the
distribution range in Galactic coordinates. The central line is $l = 143^{\circ}$, the two dotted lines are the bounds
$10^{\circ}$ from the central line on the celestial sphere, and the crosses are the candidates in SDSS DR9.  From 
this figure we can see that the width of CPS spans $\sim 20^{\circ}$, and almost all CPS candidates
rest in $130^{\circ} < l < 160^{\circ} $ and $ -25^{\circ} < b < -75^{\circ}$. CPS may extend beyond 
$l = 160^{\circ}$, this will need confirmation by another spectroscopic survey.

\par
There is no APOGEE data available with corresponding photometric and proper motion data which overlaps CPS.
\par

Because candidates of CPS from both SDSS DR9 and LAMOST DR3 are all within narrow strips in their CMD 
and $V_{gsr}$ figures, we are highly confident of membership,  their confidence levels are set to 1.

%
%
%

\subsection{Candidates of the Orphan Stream}

We select the Orphan stream candidates by the method given by \cite{new10}. In their paper, they
 defined a new coordinate system ($B_{Orphan}, \Lambda_{Orphan}$). Under this coordinate system, 
they defined a variable $B_{corr}$ to let the stream locus be at  $ B_{corr}= 0$. Here, we also use these 
symbols with the same definitions. All giant candidates 
should match the following criteria:
\par
1) Metallicities should be within  $[-2.5,  -1.6]$;
\par
2) These candidates should be within $-2^\circ < B_{corr} < 2^\circ$.
\par
3) We denote $ T_{Orphan} =  -0.0445 \Lambda^2_{Orphan} - 0.935 \Lambda_{Orphan} + 130$. 
Then their Galactic standard of rest velocities $V_{gsr}$ are within $ T_{Orphan} \pm 35$ km s$^{-1}$.
\par
4) We calculate their $g_{corr}$ by the formula: $g_{corr} = g_0 - 0.00022\Lambda^2_{Orphan}+0.034 
\Lambda_{Orphan}$.
\par
5) Proper motions in R.A. and Dec. are all less than 6 mas/yr.
\par
6)  We shift the isochrone of M92 to the place where its BHBs are at $g_0 = 17.75$. 
\par

Fig. \ref{orphan-cmd} shows the CMDs of Orphan stream candidates in SDSS and LAMOST data. The left 
panel and the right panel are the  Orphan stream CMD of  SDSS candidates and LAMOST candidates 
respectively. In each panel, the filled circles are stars in $ T_{Orphan} \pm 17.5 $km s$^{-1}$, while open 
circles are stars in $ T_{Orphan} \pm 35 $km s$^{-1}$. The isochrone of M92 is shifted to
have its BHBs at $g_0 = 17.75 $,  as \cite{new10} did.  In the LAMOST data, at least four star's photometry is
unreliable.
\par

In Fig. \ref{orphan-feh}, the metallicity distributions of Orphan stream candidates in SDSS and LAMOST data
are shown respectively by the left and right panel. In each panel, the red histogram is the metallicity distribution of stars in 
$ T_{Orphan} \pm 17.5 $km s$^{-1}$, while the black histogram is that of stars in $ T_{Orphan} \pm 35 $km s$^{-1}$. 
We find that for SDSS candidates, there are 3 metallicity peaks which are $[-2.5, -2.3]$, $[-2.2, -1.9]$, and  
$[-1.9, -1.6]$. The last two peaks are confirmed by LAMOST data. Because LAMOST metallicity is cutoff by 
$\rm [Fe/H] = -2.4$, the most metal-poor component does not appear in the LAMOST data.  If three metallicity
 peaks are real, it suggests that there are 2-3 components in the Orphan stream or that other stellar populations from other objects are overlapping in space and velocity. Fig. \ref{orphan-vgsr} 
 shows the  $V_{gsr}$  of  SDSS candidates along the  $\Lambda_{Orphan} $. The dash dot lines  are 
 $ T_{Orphan} \pm 17.5 $ km s$^{-1}$, while the dash lines are  $ T_{Orphan} \pm 35$ km s$^{-1}$. The stars with 
 metallicity in $[-2.5, -2.3]$, $[-2.2, -1.9]$, and  $[-1.9, -1.6]$ are marked by red, green and blue circles, respectively.  
 From Fig. \ref{orphan-vgsr} we can see that compared to the stars with [Fe/H] in $[-2.2, -1.9]$, $[-1.9, -1.6]$, 
 almost all stars within the most metal-poor peak have $\Lambda_{Orphan} < 0$, with only 2 stars in  
 $\Lambda_{Orphan} > 0$. This special metal pattern along the  $\Lambda_{Orphan} $ must relate to the 
 origin and evolution histories of the stream, and need further study.
\par

There is no APOGEE star candidate in the Orphan stream.
\par

We use similar formulas in Sec. \ref{sec-gd1} to calculate confidence levels. 
Let $Dv_i$ denote the difference between $V_{gsr}$ of the $i$th candidate and the trendline 
$-0.0445 \Lambda^2_{Orphan} - 0.935 \Lambda_{Orphan} + 130$ km s$^{-1}$ at its $\Lambda_{Orphan} $, 
and $\sigma^2_1 = \sum_{i\in\phi}Dv^2_i/(n-1)$, where $\phi =\{i | |Dv_i| < 3 \sigma_1 \}$ and $n$ is the 
number of the elements in $\phi$; Let $Dc_i$ denote the difference between 
 $(g-r)_0$ of the $i$th candidate and the M92 isochrone  at its $g_{corr}$, and $\sigma^2_2
 = \sum_{i\in\psi}Dc^2_i/(m-1)$, where $\psi =\{i | |Dc_i| < 3 \sigma_2 \}$ and $m$ is the number of the
 elements in $\phi$. Then we calculate the $f_i$ for the $i$th candidate by the formula: 
$f_i = (\frac{Dv_i}{\sigma_1})^2 + (\frac{Dc_i}{\sigma_2})^2$.
If $f_i \leq 2$, then the confidence level of the $i$th candidate is set  to be 1; if $ 2< f_i \leq 10$, then 
the confidence level is set to be 2; if $ f_i > 10$, then the confidence level is set to be 3.

\begin{center}

  \begin{figure}
   \includegraphics[scale=0.5]{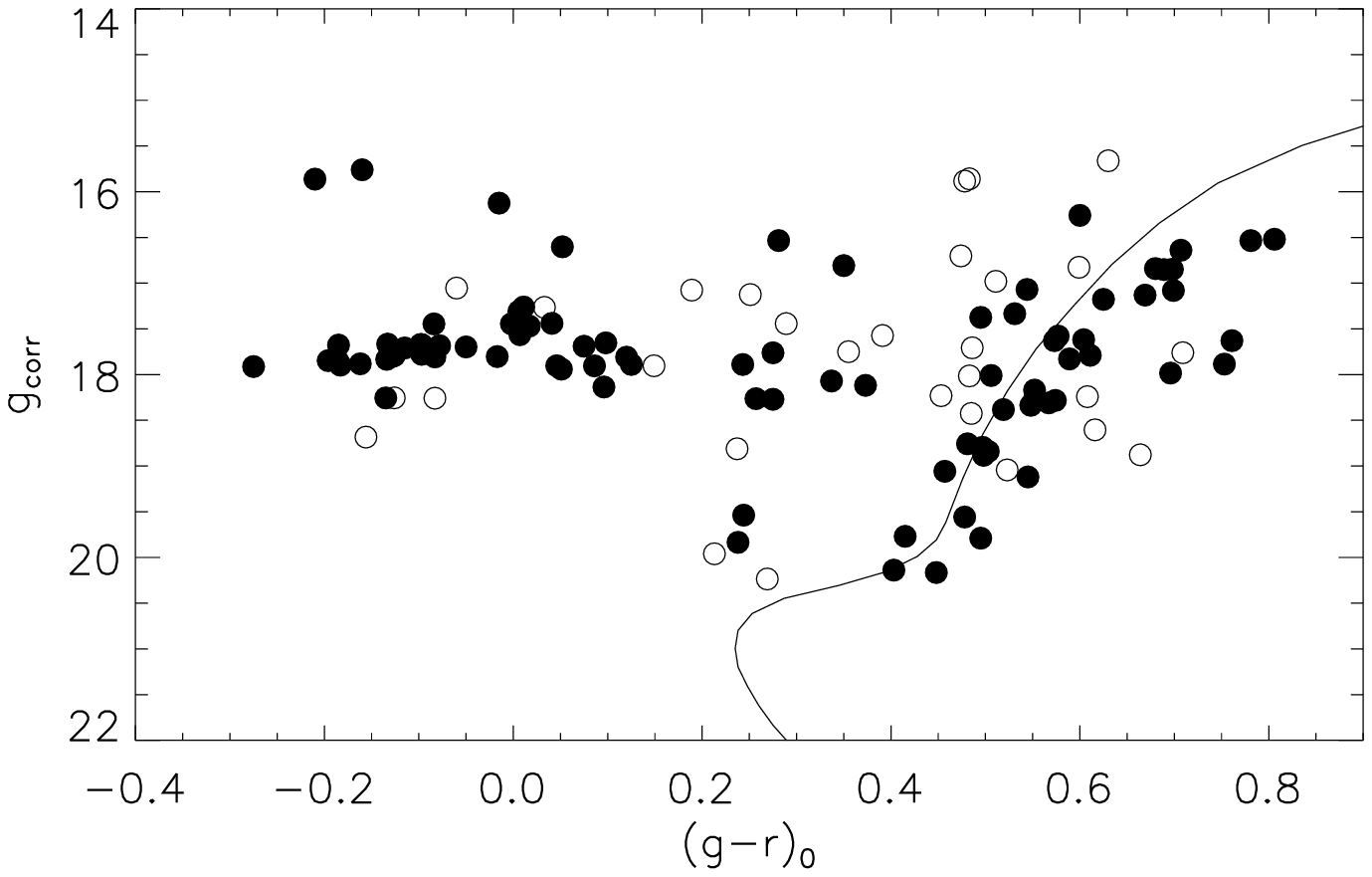} 
   \includegraphics[scale=0.5]{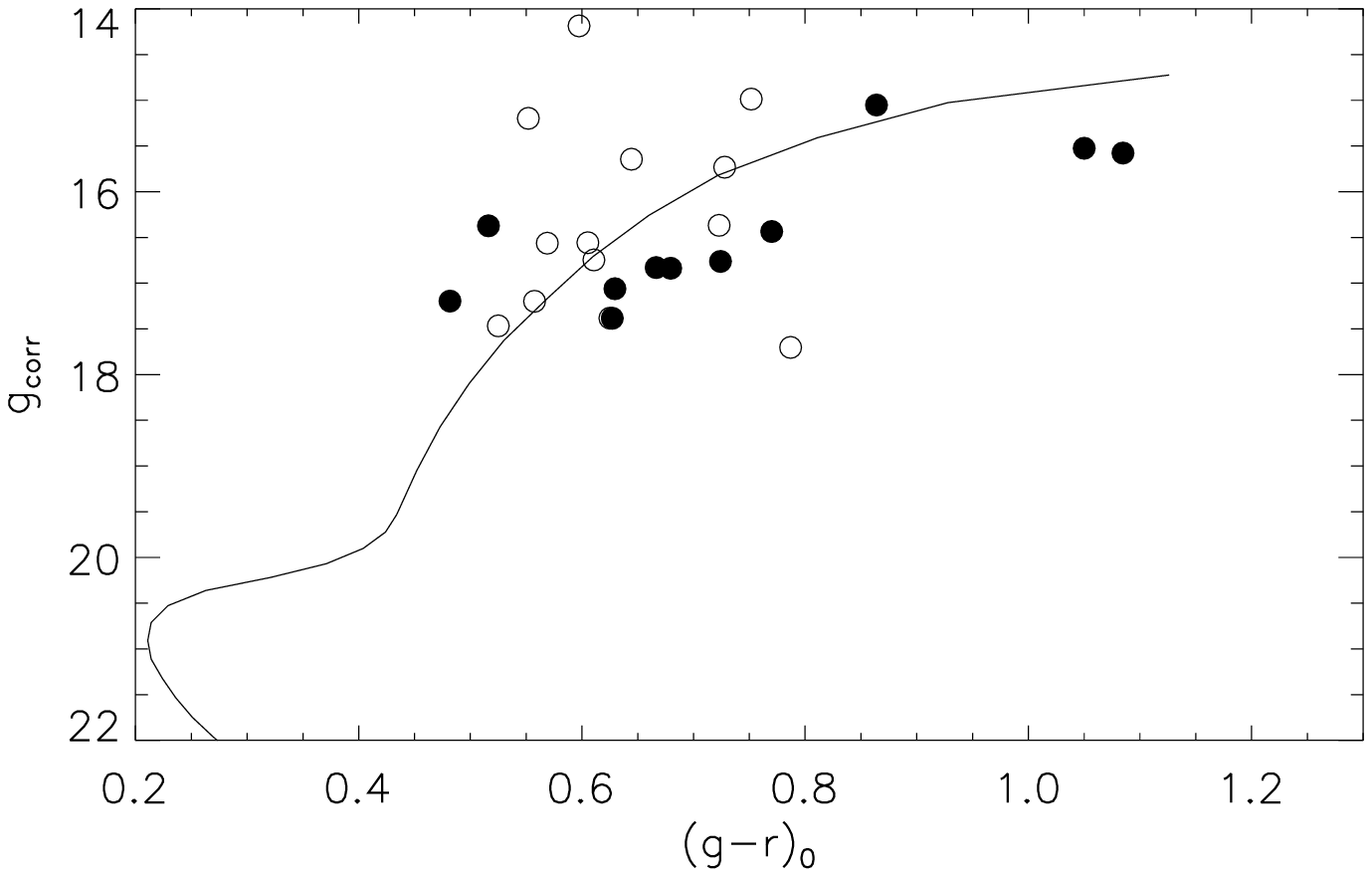}
   
      \caption{The left panel and the right panel are the CMDs of Orphan stream  candidates in SDSS 
      and LAMOST data respectively. In each panel, the filled circles are stars within $ T_{Orphan} \pm 17.5 $km 
      s$^{-1}$, while open circles are stars within $ T_{Orphan} \pm 35 $km s$^{-1}$. In each panel, the 
      isochrone of M92 is shifted to have its BHBs  at $g_0 = 17.75 $. 
      }
   \label{orphan-cmd}
   \end{figure}

 \begin{figure}   
   \includegraphics[scale=0.5]{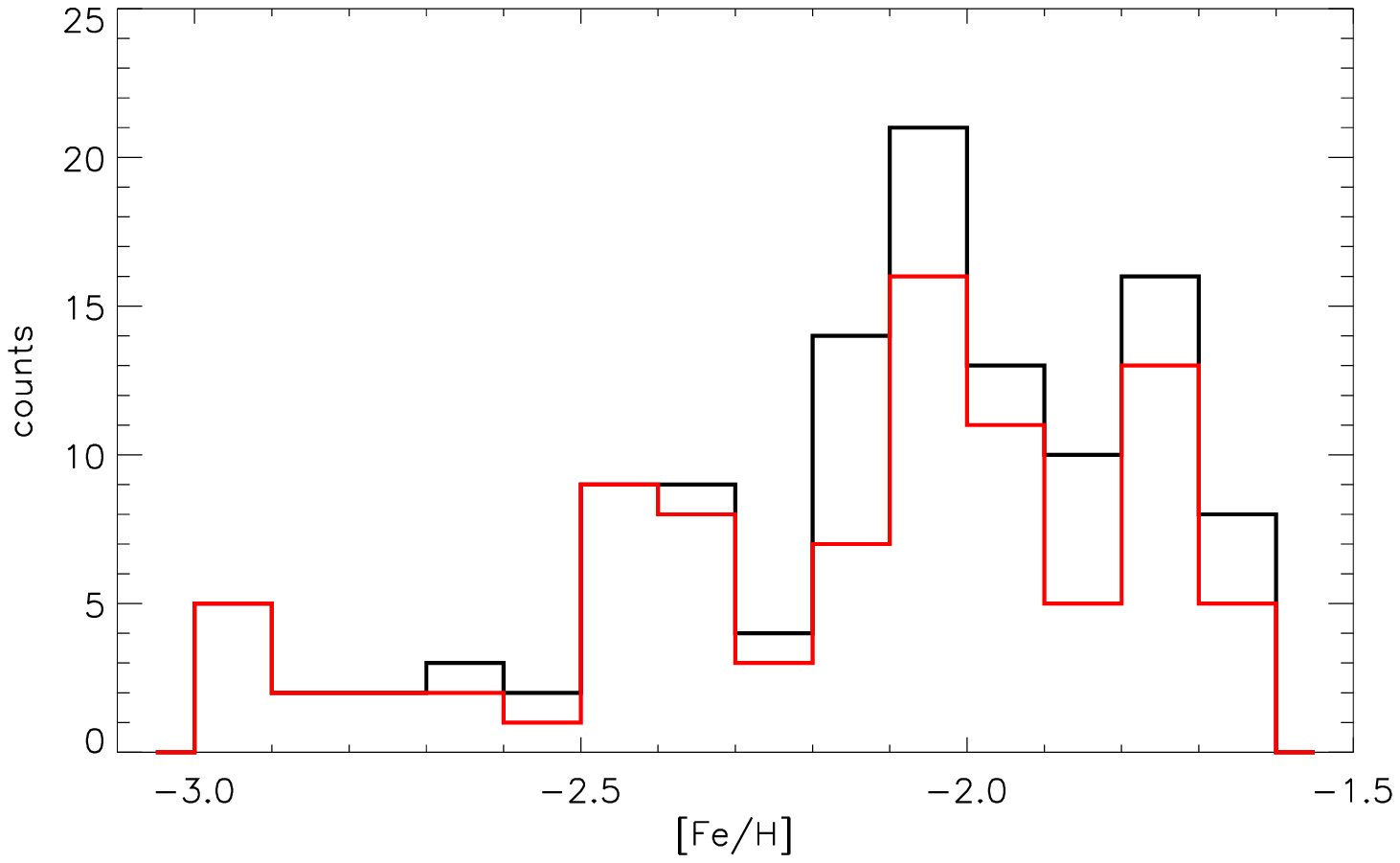} 
   \includegraphics[scale=0.5] {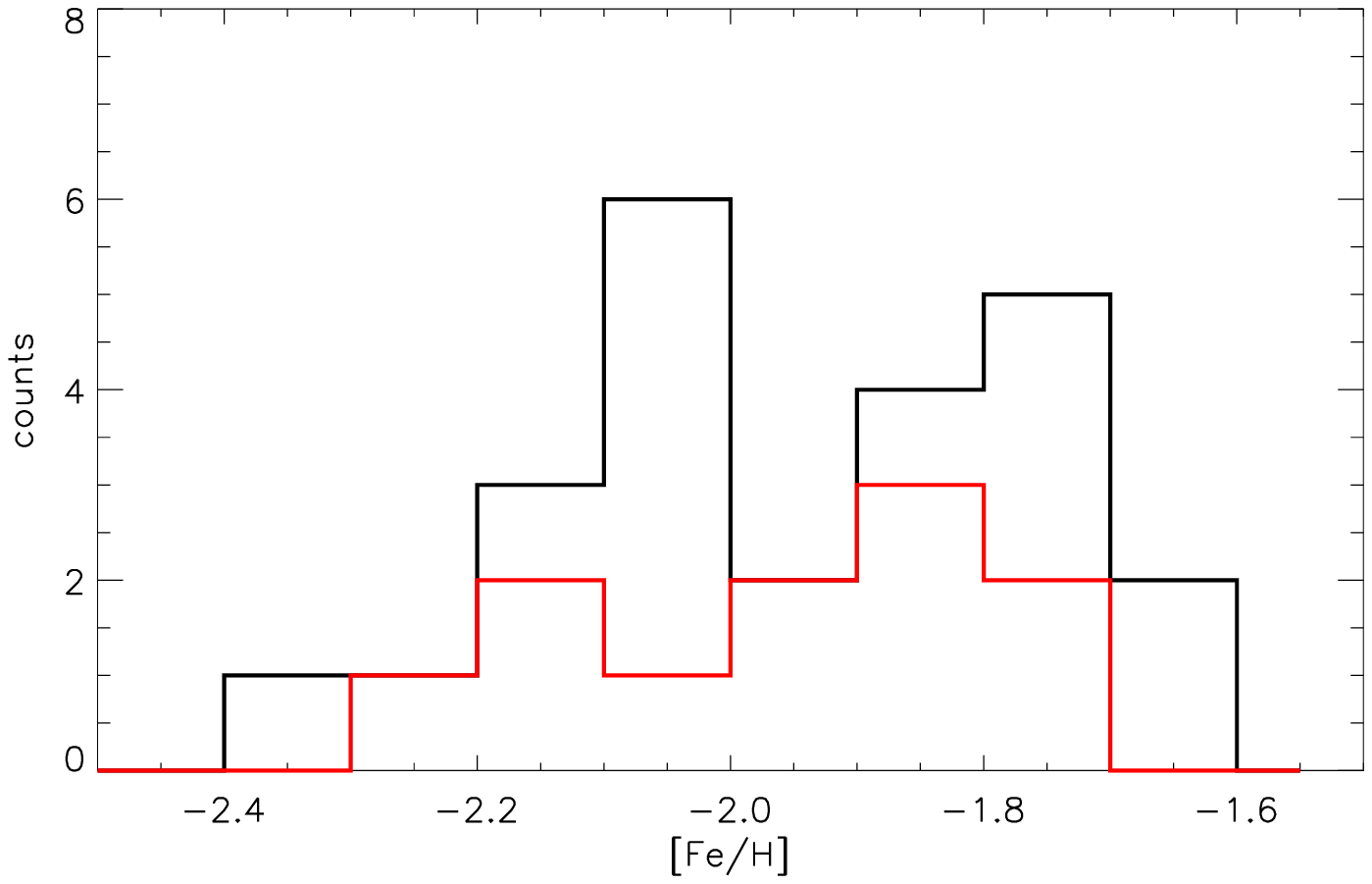}

      \caption{The left panel and the right panel are the metallicity distributions of Orphan stream candidates 
      in SDSS and LAMOST data respectively. In each panel, the red histogram is the metallicity distribution of 
      stars within $ T_{Orphan} \pm 17.5 $km s$^{-1}$, while the black histogram is for stars within $ T_{Orphan} \pm 35 $km s$^{-1}$.       }
   \label{orphan-feh}
   \end{figure}
   
   \end{center}

  \begin{figure}   
    \includegraphics{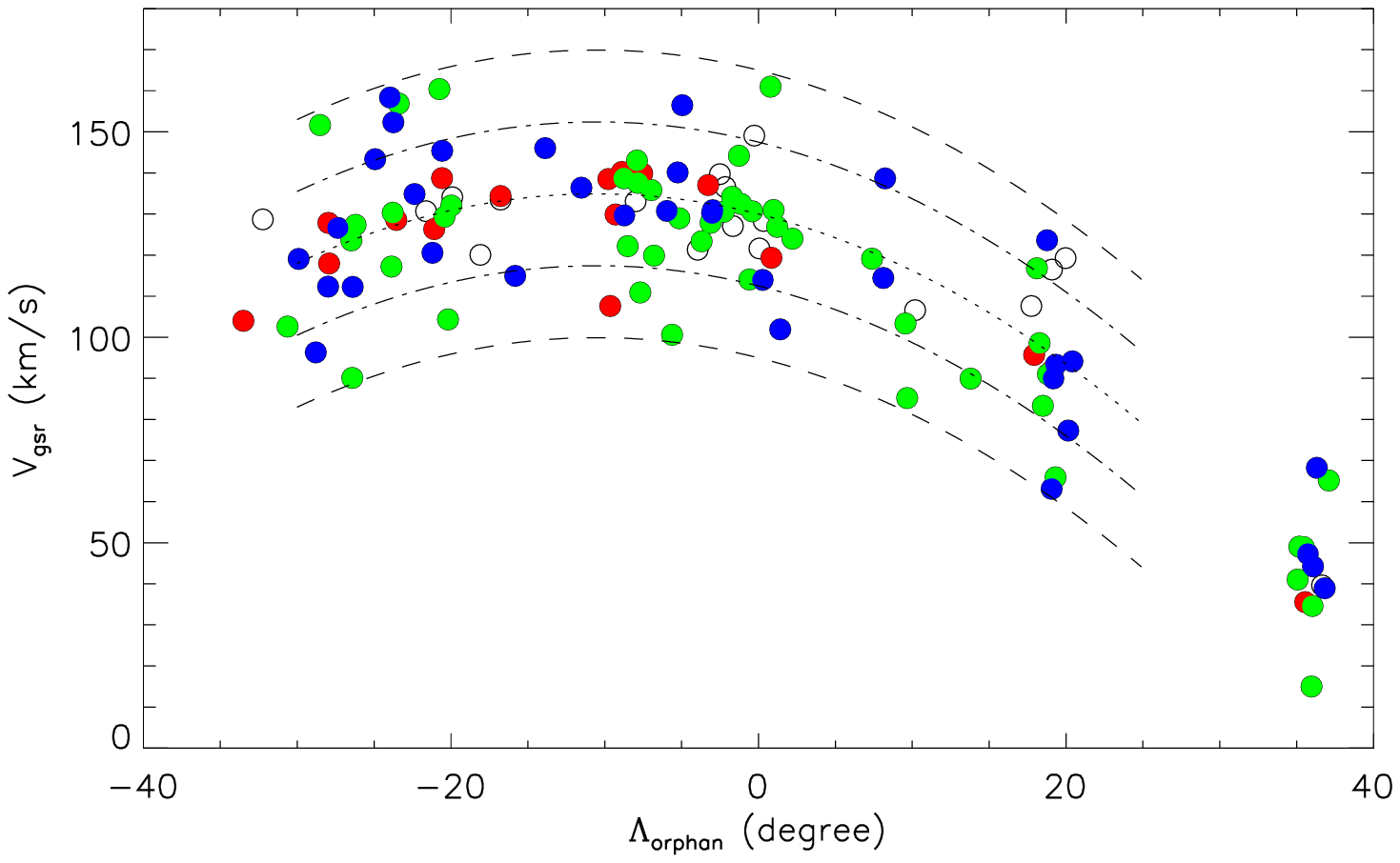} 
   
      \caption{ This figure shows the  $V_{gsr}$ of SDSS candidates along the  $\Lambda_{Orphan} $. The dash dot lines 
      are $ T_{Orphan} \pm 17.5 $ km s$^{-1}$, while the dash lines are  $ T_{Orphan} \pm 35$ km s$^{-1}$.  The central 
      dot line is $ T_{Orphan}$ km s$^{-1}$. All candidates are marked
      by circles, while stars with [Fe/H] in $[-2.5, -2.3]$, $[-2.2, -1.9]$, and  $[-1.9, -1.6]$ are marked by red, green and blue circles, respectively. }

   \label{orphan-vgsr}
   \end{figure}

\subsection{Pal 5 Tidal Stream}
We select Pal 5 tidal stream candidates following the criteria from \cite{kuz15}. All giant candidates satisfy the following 
criteria:
\par
1)Three positions $(226.3^\circ,  -2.9^\circ)$,  $(246^\circ, 7.9^\circ)$ and Pal 5 center $(229^\circ, -0.11^\circ)$ 
in (R.A., decl.)  in \cite{gri06} and the 11 centers of fields given in Table 1 by \cite{kuz15} are used to calculate the trace 
of the Pal 5 tidal stream. Firstly, these positions are classified into 2 groups: the southern group consists of the cluster center 
and positions to the south; the northern group consists of the cluster center and positions to the north. Then these positions 
in each group are converted from  (R.A., decl.) to $(l, b)$. For the trailing tail (northern part of the stream), we fit positions 
with a cubic polynomial, which is $b = 45.9816-0.0250988* l -0.0244554* l^2 + 0.000332027*l^3$. For the leading
tail (southern part of the stream),  we fit positions by a line, $b = 45.920122 -0.010919619*l$.
All candidates should be within $1^{\circ}$ of this locus.
\par
2) We use photometry of stars within $8.3'$ around Pal 5 the globular cluster center to generate a Pal 5 CMD.
\par
3) Proper motions in R.A. and Dec. are all less than 6 mas/yr.
\par
4) Metallicities are within $-2.5 < $[Fe/H] $< -0.6$. As RR Lyraes are variables, it is hard to get their
 real metallicity from one spectrum without knowing a phase. Thus, the metallicity criterion is extended to $-2$. 
 Secondly, there are very few (even no)  candidates in these 3 catalogues on the Pal 5 long trailing tail, so every 
 candidate is very valuable. We extend the metallicity criterion to -0.6. In fact, only one member candidate 
 (a red giant) has [Fe/H]$\sim -0.75$, while other red giant candidates have [Fe/H] about within $-1.4 \pm 0.4$.
\par
5) As mentioned by \cite{kuz15}, the velocity gradient along the trace is $1.0 \pm 0.1$ km s$^{-1}$deg$^{-1}$, so 
we use $RV_{corr} = RV - a$ to select candidates, where $a$ is angular distance in degree to cluster center, $a < 0$ 
for leading tail stars,  $a > 0$ for trailing tail stars, and $RV$ is the line of sight velocity.  $RV_{corr}$ should be within 
$[-70, -45]$ km s$^{-1}$.
\par
6) We use $0.00361996 g_0^3 -0.173359 g_0^2 + 2.61828 g_0  -11.471$ to fit the fiducial sequence of Pal 5 red giant 
branch given by \cite{an08}, and use the criteria of $0.00361996 g_0^3 -0.173359 g_0^2 + 2.61828 g_0  -11.471 + 
0.1 < (g - r)_0 <0.00361996 g_0^3 -0.173359 g_0^2 + 2.61828 g_0  -11.471 - 0.15$, $16 < g_0 < 19.5$ and $-2<$[Fe/H]
$< -0.6$ to select candidates in red and asymptotic giant branch. In the horizontal branch and RR Lyrae strip, we 
use the criteria of $ -0.3 < (g - r)_0 < 0.3$ and $16.7 < g_0 < 18.5$. 

\par
We find no candidate in LAMOST data, but we find 8 red giants in APOGEE. We also find 5 RR Lyraes and 
5 red giants in the SDSS DR9 spectra.
\par
Fig. \ref{pal5-cmd}, \ref{pal5-pos}, \ref{pal5-feh} and \ref{pal5-rv} are CMD, position, 
metallicity and line of sight velocity distributions of Pal 5 candidates in APOGEE and SDSS DR9, respectively.  From 
these figures we can see,  APOGEE candidates are near the tip of the red giant branch, most near the Pal 5 core; 
SDSS candidates are red giants and RR Lyraes. These SDSS red giants are far from the center while RR Lyraes 
are near the Pal 5 cluster center. Metallicities of APOGEE candidates are around $-1.1$, higher than $-1.48 \pm 0.10$ dex 
given by \cite{kuz15} and  $-1.41$ given by the Harris catalogue. 
SDSS candidates metallicity should be within $[-1.9, -1.4]$, as shown in the 
left panel of Fig. \ref{pal5-feh}, which is lower than those given by APOGEE, \cite{kuz15} and the Harris catalogue. 
For the line-of-sight velocity, its distribution of APOGEE candidates is very narrow which confirms that Pal 5 is a cool stream, though the
SDSS distribution is broader. 

As the APOGEE candidates and SDSS RR Lyraes are all in the Pal 5 cluster, so they are bona fide members, then 
their confidence levels are set to 1. The 5 SDSS red giants are all far from Pal 5 center, so their confidence levels are set to 2.

\begin{center}

  \begin{figure}
   \includegraphics[scale=0.5]{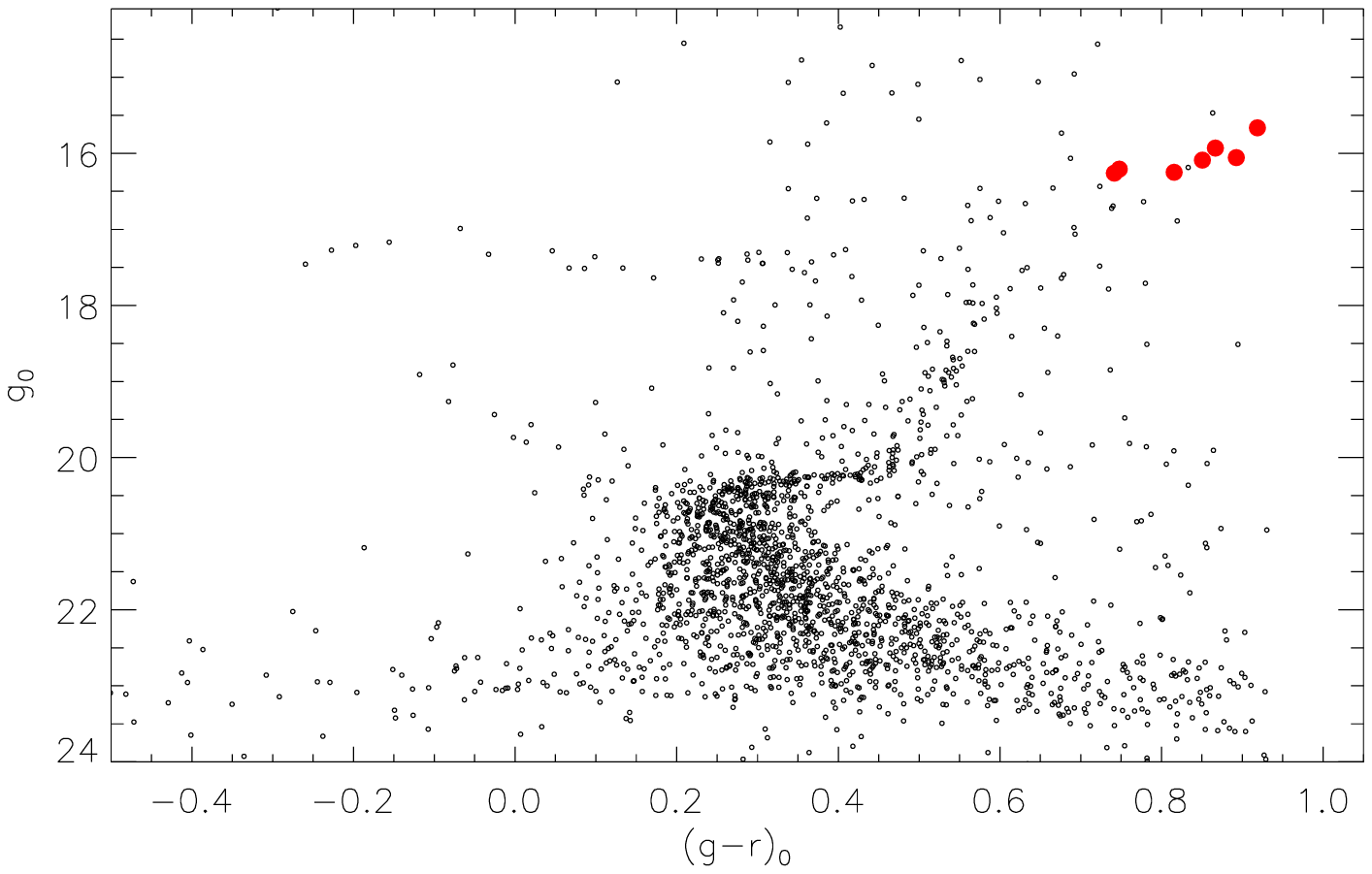}
   \includegraphics[scale=0.5] {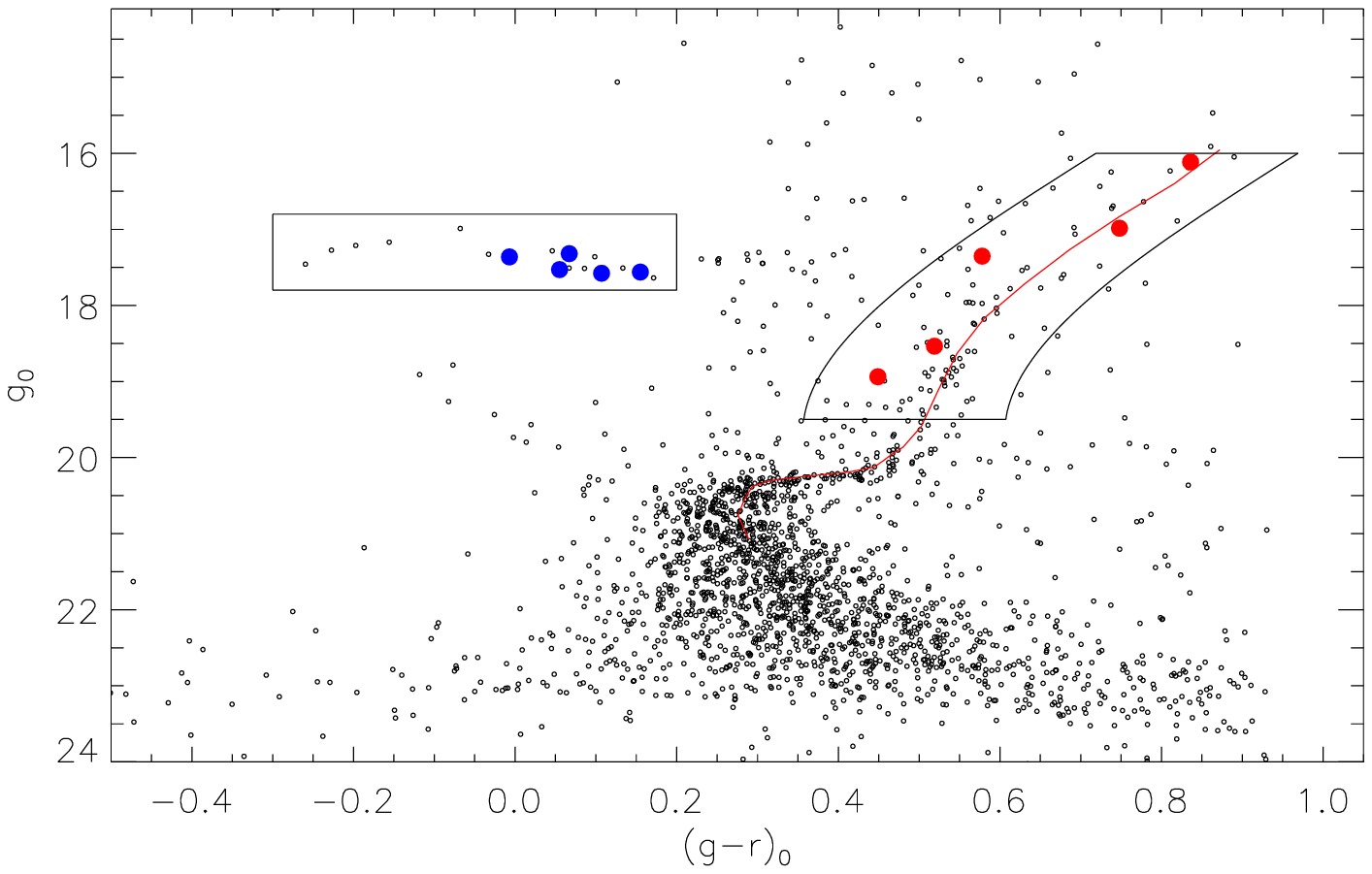}
   
      \caption{Left panel: the CMD of APOGEE candidates in the Pal5 tidal stream; Right  panel: the CMD of SDSS candidates 
      in the Pal5 tidal stream. In each panel, the background stars are within $8.3'$ around Pal 5 cluster center. Red circles are red giants,
      while blue circles are RR Lyraes. In the right panel, areas enclosed by black lines are where we select candidates in SDSS DR9, 
      while the red line is the fiducial sequence of Pal 5 given by \cite{an08}.}
   \label{pal5-cmd}
   \end{figure}

 \begin{figure}   
   \includegraphics[scale=0.5]{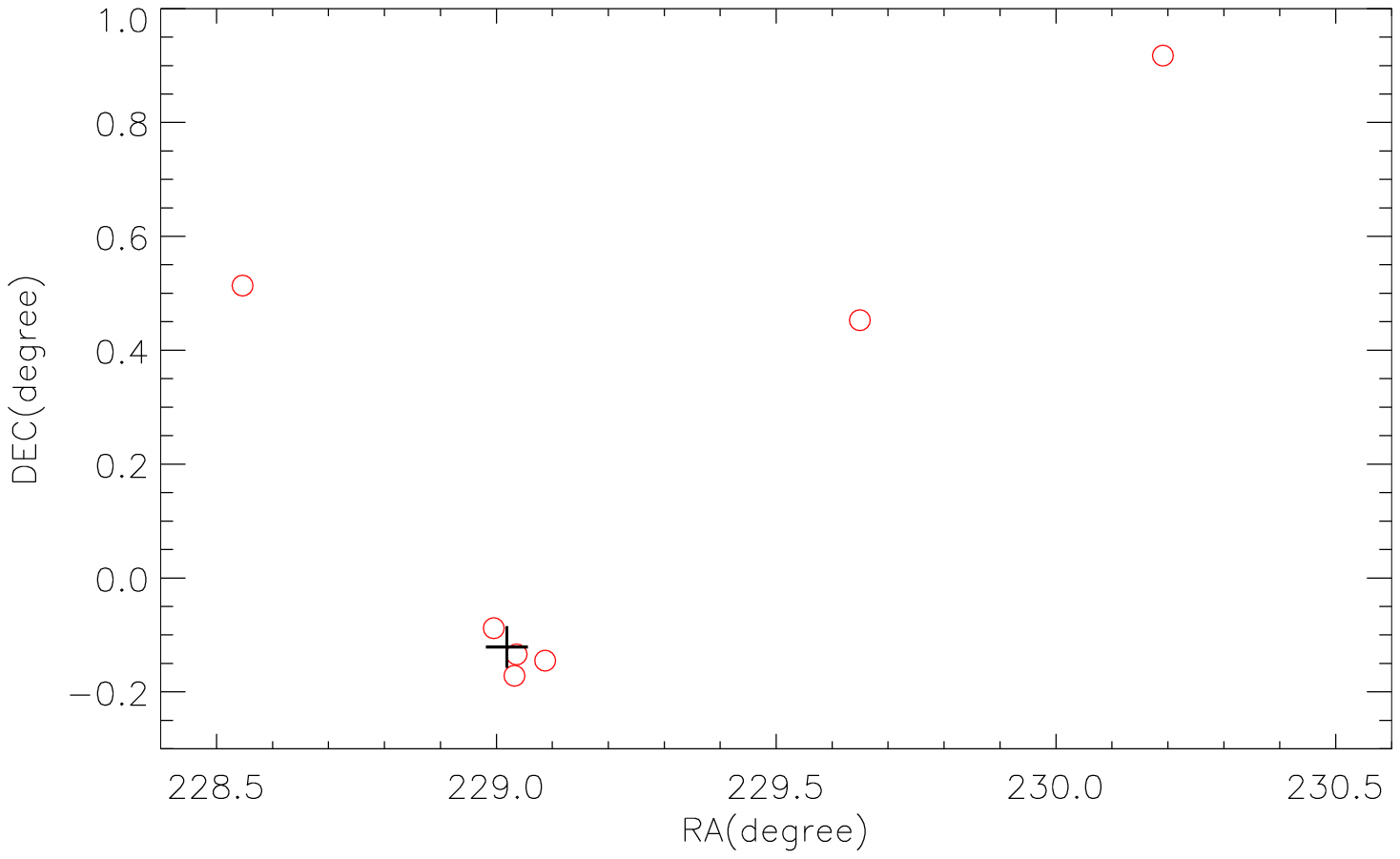} 
   \includegraphics[scale=0.5]{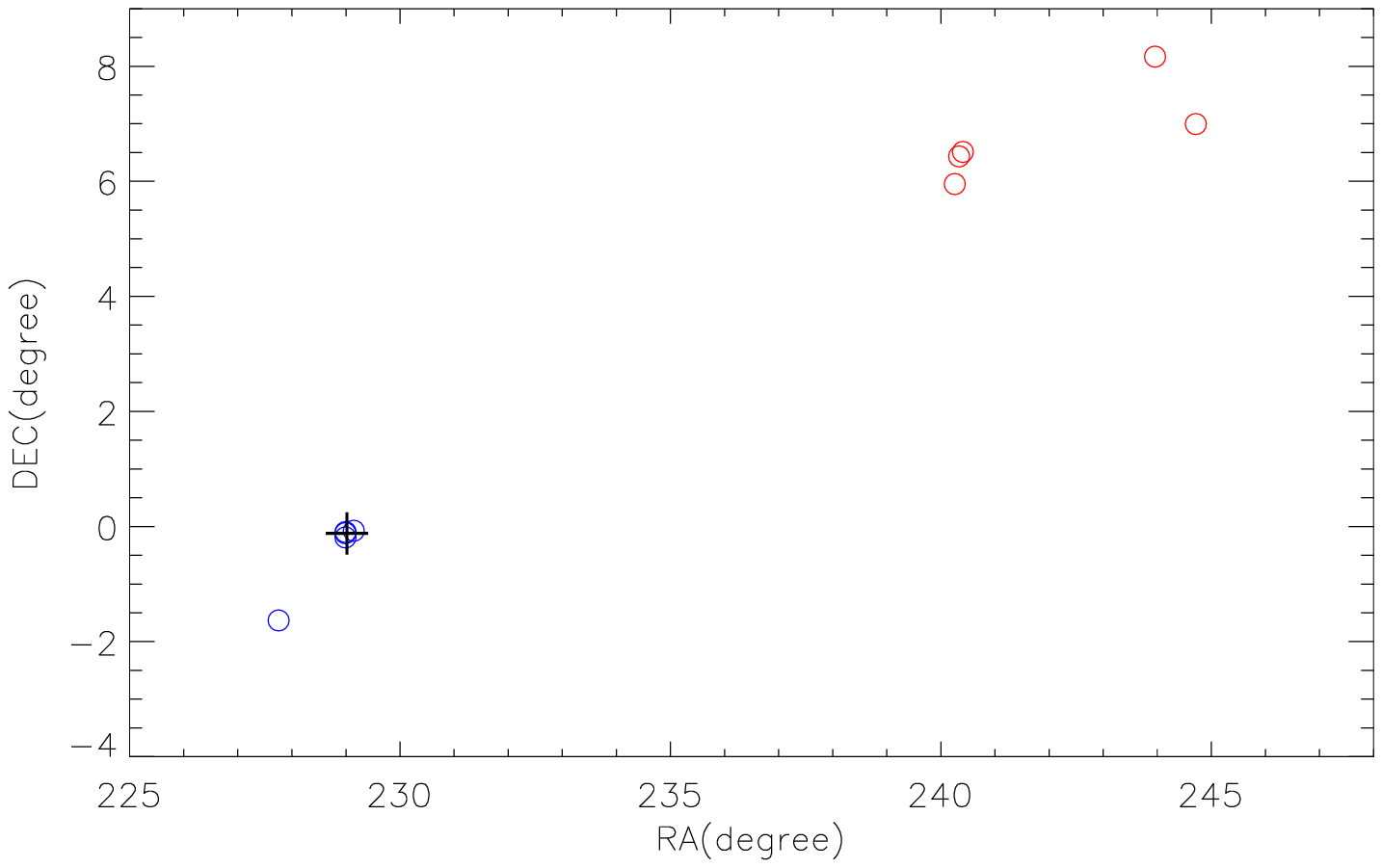}
   
      \caption{ Left panel: APOGEE candidate positions; Right panel: SDSS candidate positions.  
      Red circles are red giants, while blue circles are RR Lyres. Black cross is Pal 5 cluster center in each panel.}
   \label{pal5-pos}
   \end{figure}
 
  \begin{figure}   
   \includegraphics[scale=0.5]{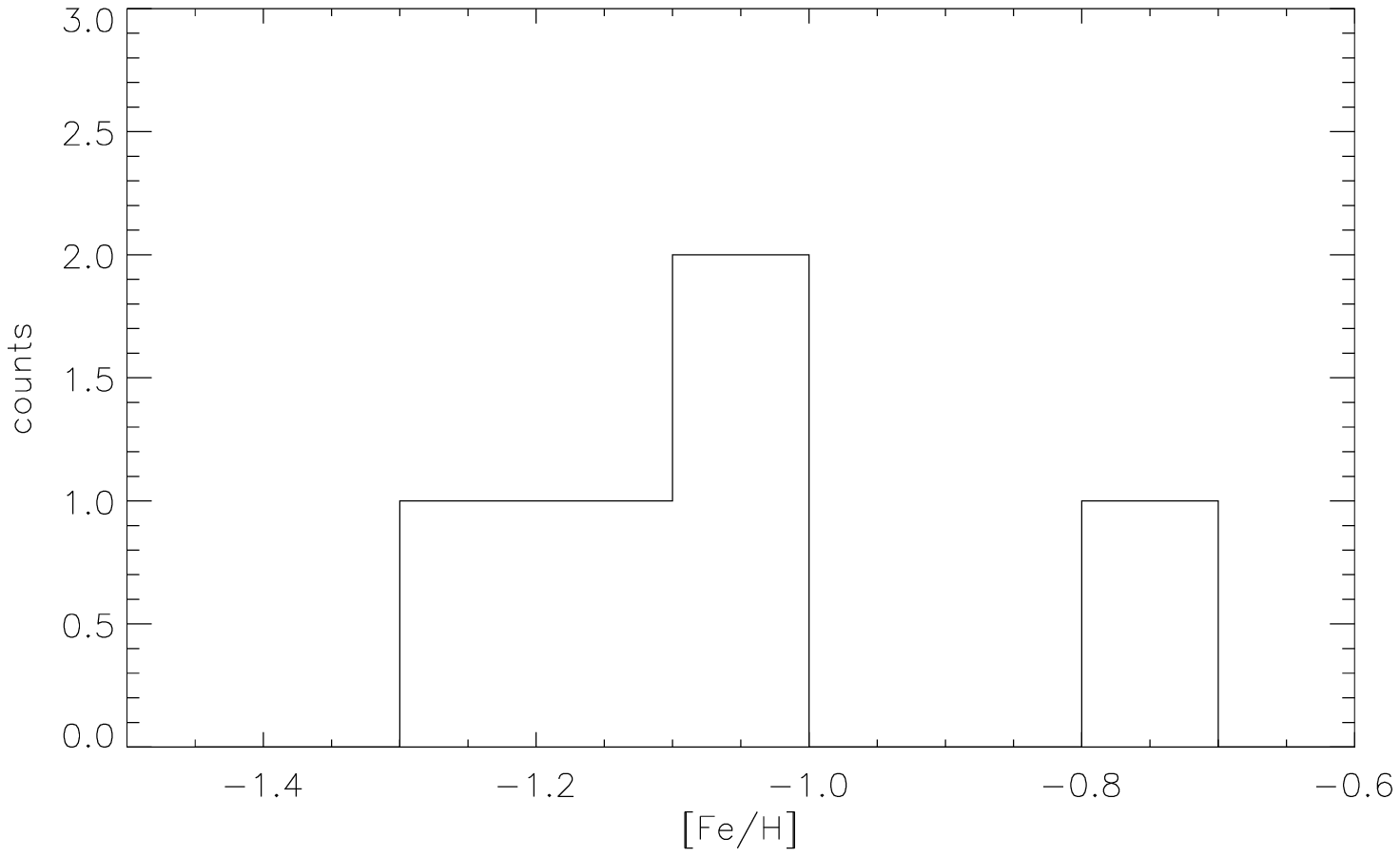} 
   \includegraphics [scale=0.5]{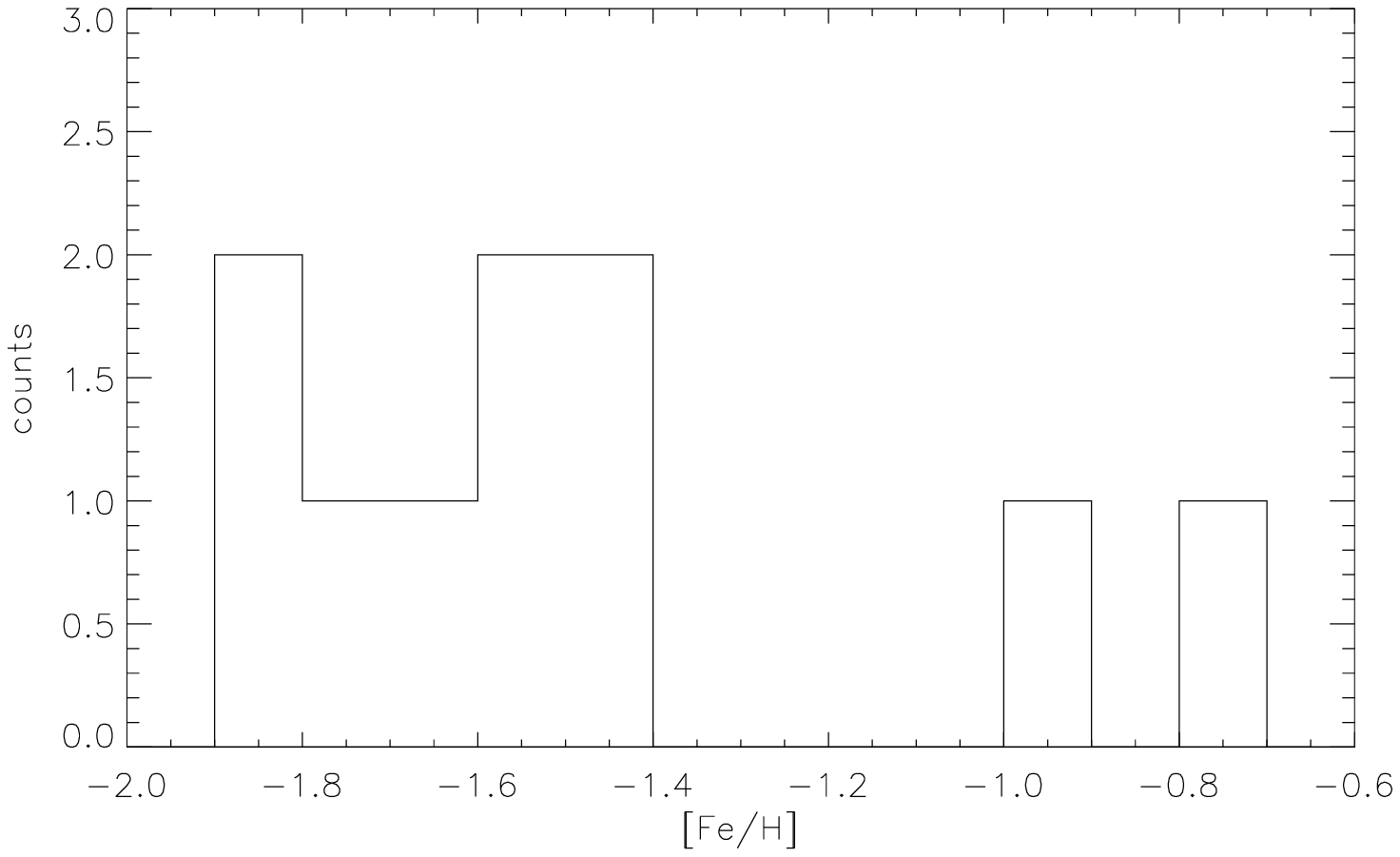}
   
      \caption{ Left panel: Metallicity distribution of Pal 5 candidates in APOGEE; Right panel:  Metallicity distribution
      of Pal 5 candidates in SDSS DR9. }
   \label{pal5-feh}
   \end{figure}
   
   \begin{figure}   
   \includegraphics[scale=0.5]{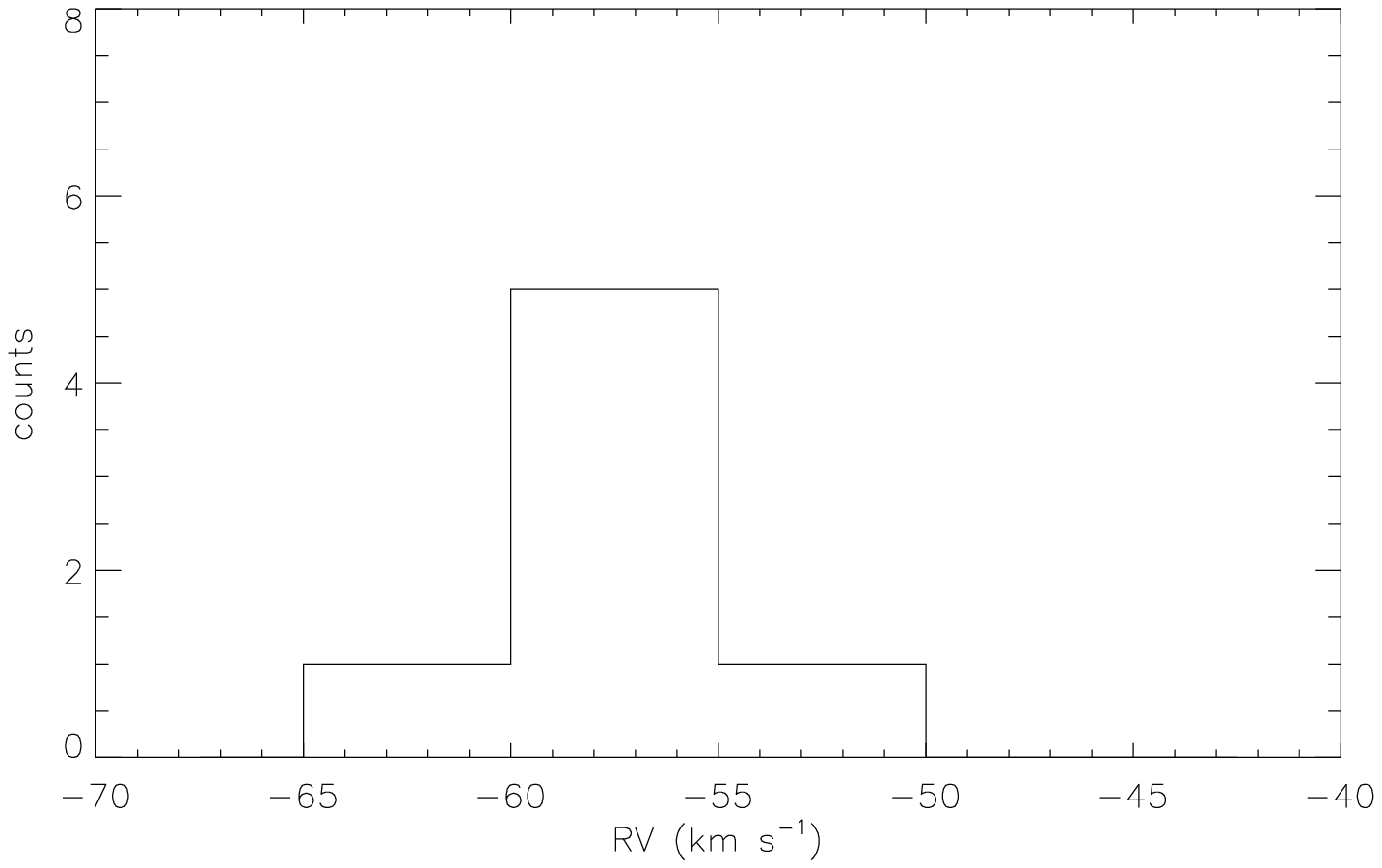} 
   \includegraphics [scale=0.5]{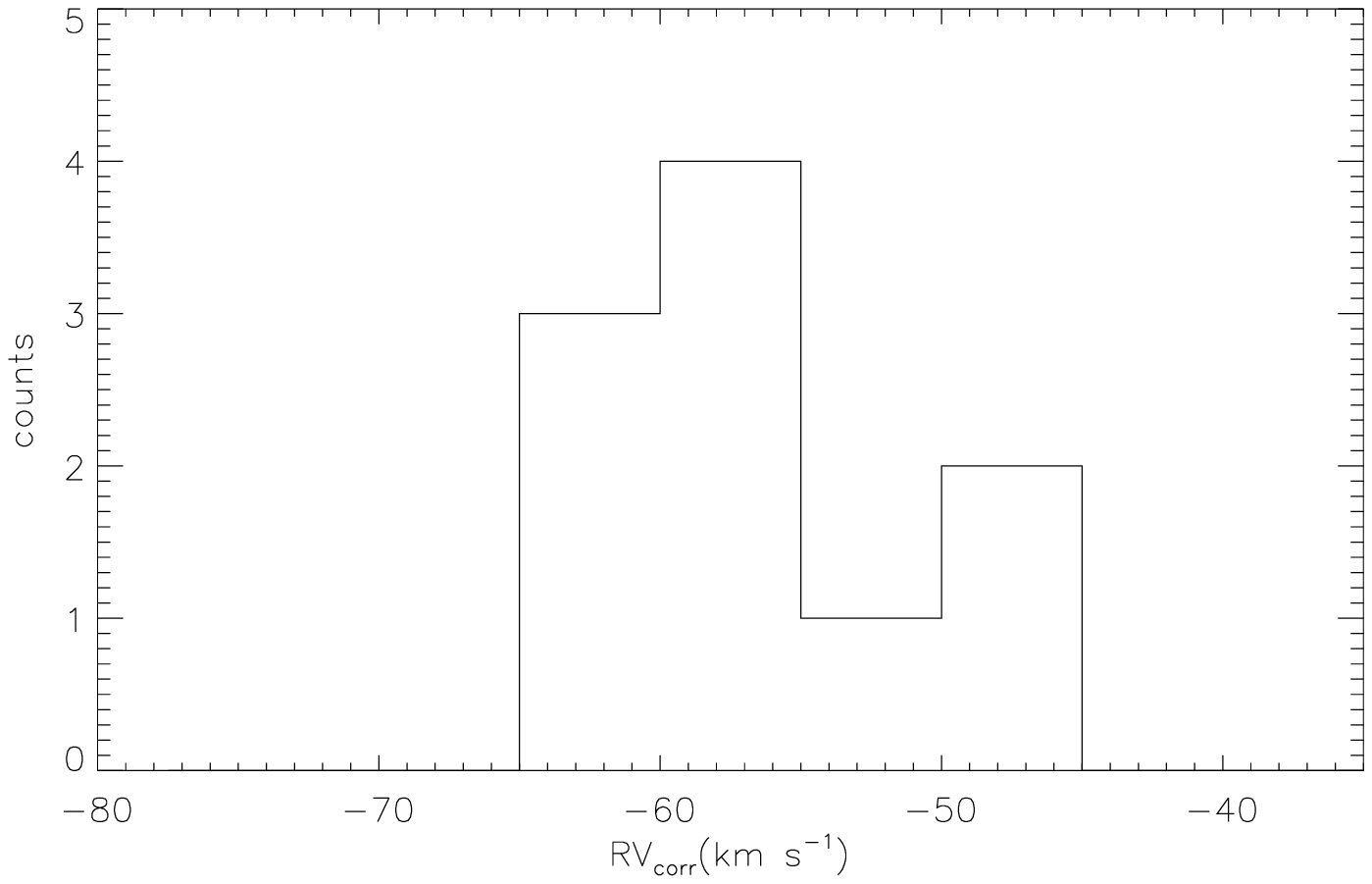}
   
      \caption{ Left panel: Line of sight velocity distribution of APOGEE candidates. Because APOGEE candidates 
      all are around Pal 5 cluster center, corrected velocity is not used.  Right panel: Corrected velocity distribution 
      of SDSS candidates. }
   \label{pal5-rv}
   \end{figure}

\end{center}

\section{Conclusion and Discussion}
In this paper, we present 3 tables of high confidence candidate stellar members of the GD-1, CPS, Orphan 
and Pal 5 tidal streams from LAMOST DR3, SDSS DR9 and APOGEE spectroscopic catalogs. 
In LAMOST DR3, we find 20, 4, 24 high confidence candidates of the GD-1 stream, CPS and the Orphan stream, respectively.  
In SDSS DR9, we find 59, 118, 10  high confidence candidates of CPS, Orphan stream and Pal 5 tidal stream, respectively.
In APOGEE, we find 7 Pal 5 high confidence candidates. 
Table \ref{can-lamost} lists the LAMOST DR3 candidates, including ID, position, spectral type, 
and radial velocity, [Fe/H], log $g$, $T_{\textrm{eff}}$ with their errors. Table \ref{can-SDSS} lists the SDSS data,
including plate, mjd, fiberid, position, spectral type, and radial velocity, [Fe/H], FEHWBG, log $g$, $T_{\textrm{eff}}$ with their errors. Table \ref{can-apo}
gives the information for each APOGEE candidate, including ID, position, and  radial velocity, $T_{\textrm{eff}}$, [Fe/H], log $g$ with their errors. 
The last column in these 3 tables show the confidence level (discribed by 1, 2, 3) of each candidates, with a higer number indicating lower
confidence.

\par
Of note:
(1) The brightest stars of GD-1 and Orphan streams, are all from LAMOST data, so the LAMOST data supplements the bright end
of these streams.

(2) LAMOST and SDSS Orphan stream data show that there may be 2 or 3
metallicty peaks, and the most metal-poor peak rests at $\Lambda_{Orphan}<0 $. 
Alternatively, there may be stars from other streams or coherent background or foreground halo structures in
this direction on the sky. The Orphan stream may span a broader area for regions beyond $\Lambda_{Orphan} < -20^{\circ}$ 
and $\Lambda_{Orphan} > 10^{\circ}$.

(3) The cataloged APOGEE metallicity for Pal 5 is around $-1.2$ which is significantly higher 
than that given by the globular cluster and SDSS literature which quote $\rm [Fe/H] \sim -1.4$.
APOGEE spectra are obtained with a high resolution infrared spectrograph and measure elements besides
Iron or Magnesium or Calcium to determine metallicity.

\par

In the future, we plan to continue to probe the abundances of stream stars, including searches for gradients in
abundance along the streams, in order to better understand the streams' formation and evolution histories.

Finally, we hope to use these additional candidate stream stars' velocity, metallicity and membership
information to improve the stream orbits.   A fit of the four orbits simultaneously to a single Milky Way 
potential can then be used to constrain the potential significantly better than fitting any single 
stream by itself and help resolve remaining questions about the extent and shape of our
dark matter halo \cite{new10,wil09,lmj09}.

\begin{acknowledgements}

This research is supported by the Natural Science Foundation of China for the Youth under grants Y011161001, National Science Foundation of China (NSFC) under grant No. 11403056, the National Natural Science Foundation of China (11673036) and NSFC Key
Program NSFC-11333004. Support was provided by the US NSF LAMOST-PLUS grant.  Y. Wu acknowledges the fund supplied by the Guangdong Provincial Engineering Technology Research Center for Data Science. We also thank the referee for valuable advice. 
\par

Guoshoujing Telescope (the Large Sky Area Multi-Object Fiber Spectroscopic Telescope LAMOST) is a National Major Scientific Project built by the Chinese Academy of Sciences. Funding for the project has been provided by the National Development and Reform Commission. LAMOST is operated and managed by the National Astronomical Observatories, Chinese Academy of Sciences.
\end{acknowledgements}

\begin{landscape}

\begin{table}\centering
\begin{threeparttable}
\caption{Candidates in LAMOST}
\label{can-lamost}
\begin{tabular}{cccccccccccccc}\hline   
stream & obsid\tnote{a} &   subclass & ra    &  dec           &   $rv$\tnote{b}         &   $T_{\textrm{eff}}$  &   [Fe/H]    &   log$g$   &    $rv$\_err & $T_{\textrm{eff}}$\_err & [Fe/H]\_err &  log$g$\_err & level\tnote{b}\\
         &                     &         &      (deg)          & (deg)                & (km s$^{-1}$)& (K)  &             &             & (km s$^{-1}$) & (K)          &                   &          &     \\\hline\noalign{\smallskip}
GD1 &  196716136  & G5  & 125.22538757  &  -2.78869820  &  253.37  & 5001.74  &   -1.78  &    2.32  &   29.10  &  186.84  &    0.29  &    1.08 & 3\\ 
GD1 &  187405120  & G7  & 137.72410583  &  21.93488503  &  135.74  & 4459.66  &   -1.55  &    1.41  &   48.83  &  344.33  &    0.61  &    1.21 & 3\\
GD1 &  132513055  & G0  & 139.29585266  &  23.10439491  &  119.02  & 5212.79  &   -1.86  &    2.86  &   42.48  &  211.86  &    0.37  &    1.16 & 3\\
GD1 &  313108245  & K3  & 125.95130157  &  -1.41831505  &  261.85  & 4246.94  &   -1.74  &    0.55  &   17.33  &   64.31  &    0.09  &    0.29 & 3\\
GD1 &  343405145  & G3  & 155.50750732  &  41.43789673  &  -75.93  & 4489.56  &   -1.98  &    0.88  &   20.57  &   61.48  &    0.10  &    0.37& 1\\
GD1 &  136814012  & F2  & 153.97523499  &  40.22281265  &  -29.44  & 5034.87  &   -2.21  &    2.36  &   48.94  &  266.52  &    0.47  &    1.21& 2\\
GD1 &   31709120  & G3  & 153.46257019  &  41.59279633  &  -84.04  & 4815.06  &   -1.64  &    1.86  &   25.35  &  250.98  &    0.41  &    1.10& 2\\
GD1 &   19114174  & G0  & 143.17625427  &  28.68422318  &   33.63  & 4878.86  &   -2.24  &    1.86  &   47.26  &  216.92  &    0.36  &    1.08& 1\\
 $\cdots$ & $\cdots\cdots$\\
 \noalign{\smallskip}\hline

\end{tabular}
 \begin{tablenotes}
        \footnotesize
        \item[a] Observation ID: the same object in different observation has different obsid which is unique for all LAMOST spectra.
        \item[b] Heliocentric heliocentric velocity.
        \item[c] The confidence level: 1 represents highest confidence. 
 \end{tablenotes}
      
Note: Only a portion of Table is shown here for illustration. The whole Table
contains information of 48 stream candidates in LAMOST is available in the online electronic version.

\end{threeparttable}
\end{table}

\begin{table}\centering
\begin{threeparttable}
\caption{Candidates in SDSS}
\label{can-SDSS}

\begin{tabular}{cccccccccccccccccc}\hline   

stream &  plate &mjd &fiberid &   subclass &   ra &  dec & $rv$\tnote{a} &   $T_{\textrm{eff}}$ &   [Fe/H] & FEHWBG &log$g$ &  $rv$\_err &$T_{\textrm{eff}}$ \_err & [Fe/H]\_err& FEHWBG\_err &log$g$\_err& level\tnote{b} \\
            &           &.     &           &                  &  (deg) & (deg) & (km s$^{-1}$)& (K) &  &     &        & (km s$^{-1}$)&(K)          &                   &                         &       &\\\hline\noalign{\smallskip}
 Cetus &   2864  &  54467  &     91  & F9  &  18.992146  & -10.615119  &  -61.74  & 4868.47  &   -2.16  &  &    1.62  &    1.62  &   55.22  &    0.06  &    &    0.14 & 1\\
 Cetus &   2878  &  54465  &     59  &A0p  &  22.955864  & -10.540712  &  -51.03  & 8456.75  &   -2.10  &   -2.10  &    3.39  &    7.31  &  141.34  &    0.29  &    0.29  &    0.29 & 1\\
 Cetus &   2864  &  54467  &     49  & A0  &  19.162479  & -10.344024  &  -54.14  & 7878.93  &   -1.72  &   -2.46  &    3.03  &    4.75  &  102.53  &    0.06  &    0.14  &    0.27 & 1\\
 Cetus &   3109  &  54833  &    111  & F9  &  16.801901  & -10.321400  &  -60.38  & 4473.98  &   -2.28  &    &    1.26  &    1.53  &  151.87  &    0.03  &    &    0.13 & 1\\
 Cetus &   2864  &  54467  &    102  & G2  &  19.334705  & -10.221816  &  -68.46  & 5151.40  &   -2.13  &    &    1.97  &    3.39  &   45.32  &    0.03  &    &    0.11 & 1\\
 Cetus &   2865  &  54503  &    269  & G2  &  24.394780  &  -9.886655  &  -65.46  & 5153.11  &   -2.40  &    &    2.24  &    3.58  &   59.23  &    0.05  &    &    0.22 & 1\\
 Cetus &   2878  &  54465  &    312  &A0p  &  20.752808  &  -9.660568  &  -73.00  & 8294.23  &   -1.92  &   -2.42  &    3.25  &    8.20  &  144.40  &    0.16  &    0.28  &    0.07 & 1\\
 Cetus &   2864  &  54467  &    351  & F9  &  17.415293  &  -9.638273  &  -70.09  & 4912.60  &   -2.25  &    &    1.48  &    2.09  &    9.19  &    0.05  &    &    0.12 & 1\\
 $\cdots$ & $\cdots\cdots$\\\hline
\noalign{\smallskip}

\end{tabular}
 \begin{tablenotes}
        \footnotesize
        \item[a] Heliocentric heliocentric velocity.
        \item[b] The confidence level: the lower number the higher confidence. 
 \end{tablenotes}
      
Note: Only a portion of Table is shown here for illustration. The whole Table
contains information of 187 stream candidates in SDSS is available in the online electronic version.

\end{threeparttable}
\end{table}

\begin{table}\centering
\begin{threeparttable}

\caption{Candidates in APOGEE}
\label{can-apo}
\begin{tabular}{ccccccccccccc}\hline
  
  stream& ID&   ra&  dec&   $rv$\tnote{a}&  $T_{\textrm{eff}}$&   [Fe/H]&   log$g$&    $rv$\_err&   $T_{\textrm{eff}}$\_err&   [Fe/H]\_err&   log$g$\_err& level\tnote{b} \\ 
            &       &(deg)&(deg)&(km s$^{-1}$)&(K)&.         &         & (km s$^{-1}$)&.   (K)        &                       &                       &                \\\hline\noalign{\smallskip}
    Pal5&  2M15160773-0010183  & 229.032225  &  -0.171773  &  -58.11  & 5119.35  &   -0.79  &    1.98  &    0.38  &   91.47  &    0.05  &    0.11 &1\\
    Pal5&   2M15162090-0008426  & 229.087113  &  -0.145193  &  -57.70  & 4408.65  &   -1.09  &    1.12  &    0.16  &   91.47  &    0.05  &    0.11 &1\\
    Pal5&   2M15160866-0008031  & 229.036088  &  -0.134211  &  -64.40  &          &          &          &    0.30  &         &          &         &1\\
    Pal5&   2M15155888-0005171  & 228.995349  &  -0.088101  &  -56.77  & 4400.08  &   -1.25  &    0.87  &    0.12  &   91.47  &    0.05  &    0.11 &1\\
    Pal5&   2M15183589+0027100  & 229.649545  &   0.452789  &  -57.41  & 4584.16  &   -1.07  &    1.94  &    0.10  &   91.47  &    0.05  &    0.11 &1\\
    Pal5&   2M15141111+0030487  & 228.546300  &   0.513550  &  -54.43  &          &          &          &    4.27  &         &          &        &1\\
    Pal5&   2M15204588+0055032  & 230.191207  &   0.917572  &  -56.09  & 4494.36  &   -1.18  &    1.27  &    0.09  &   91.47  &    0.05  &    0.11 &1\\

\noalign{\smallskip}\hline

\end{tabular}

 \begin{tablenotes}
        \footnotesize
        \item[a] Heliocentric heliocentric velocity.
        \item[b] The confidence level: 1 indicates highest confidence. 
 \end{tablenotes}

\end{threeparttable}
\end{table}

\end{landscape}

\label{lastpage}
\end{document}